\documentclass[12pt,english]{article}
\usepackage[T1]{fontenc}
\usepackage[latin9]{inputenc}
\usepackage{geometry}
\usepackage{babel}
\usepackage{amsmath}
\usepackage{amsthm}
\usepackage{amssymb}
\usepackage{graphicx}
\usepackage{setspace}
\usepackage[authoryear]{natbib}
\usepackage[unicode=true,pdfusetitle,
 bookmarks=true,bookmarksnumbered=false,bookmarksopen=false,
 breaklinks=false,pdfborder={0 0 1},backref=false,colorlinks=false]
 {hyperref}
\hypersetup{
 colorlinks=true,citecolor=blue, urlcolor=cyan, linkcolor=black}

\makeatletter

\providecommand{\tabularnewline}{\\}

\theoremstyle{plain}
    \ifx\thechapter\undefined
      \newtheorem{assumption}{\protect\assumptionname}
    \else
      \newtheorem{assumption}{\protect\assumptionname}[chapter]
    \fi
\theoremstyle{plain}
    \ifx\thechapter\undefined
	    \newtheorem{thm}{\protect\theoremname}
	  \else
      \newtheorem{thm}{\protect\theoremname}[chapter]
    \fi

\usepackage{fontawesome}

\makeatother

\providecommand{\assumptionname}{Assumption}
\providecommand{\theoremname}{Theorem}

\begin{document}
\date{}
\title{{\huge{}Conditional inference in }\textsl{\huge{}cis}{\huge{}-Mendelian
randomization using weak genetic factors}\textsc{\huge{}}\thanks{We thank Jack Bowden, Francis DiTraglia, Apostolos Gkatzionis, Alexei
Onatski, Richard Smith, and Chris Wallace for helpful discussions.
\protect \\
E-mail: Ashish Patel (\protect\href{mailto:ashish.patel@mrc-bsu.cam.ac.uk}{ashish.patel@mrc-bsu.cam.ac.uk});
Dipender Gill (\protect\href{mailto:dipender.gill@imperial.ac.uk}{dipender.gill@imperial.ac.uk});
Paul Newcombe (\protect\href{mailto:paul.newcombe@astrazeneca.com}{paul.newcombe@astrazeneca.com});
Stephen Burgess (\protect\href{http://sb452@medschl.cam.ac.uk}{sb452@medschl.cam.ac.uk})}}
\author{Ashish Patel{\small{} }\textsuperscript{{\small{}a}}{\small{},}
Dipender Gill{\small{} }\textsuperscript{{\small{}b}}, Paul Newcombe{\small{}
}\textsuperscript{{\small{}a}}, \& Stephen Burgess{\small{} }\textsuperscript{{\small{}a,c}}}
\maketitle
\begin{center}
{\small{}\vskip -2em}{\footnotesize{}}\textsuperscript{{\footnotesize{}a}}{\footnotesize{}
}{\small{}MRC Biostatistics Unit, University of Cambridge}\\
{\footnotesize{}}\textsuperscript{{\footnotesize{}b}}{\footnotesize{}
}{\small{}Department of Epidemiology and Biostatistics, Imperial College
London}\\
{\footnotesize{}}\textsuperscript{{\footnotesize{}c}}{\footnotesize{}
}{\small{}Cardiovascular Epidemiology Unit, University of Cambridge}{\small\par}
\par\end{center}
\begin{abstract}
Mendelian randomization is a widely-used method to estimate the unconfounded
effect of an exposure on an outcome by using genetic variants as instrumental
variables. Mendelian randomization analyses which use variants from
a single genetic region (\textsl{cis}-MR) have gained popularity for
being an economical way to provide supporting evidence for drug target
validation. This paper proposes methods for \textsl{cis}-MR inference
which use the explanatory power of many correlated variants to make
valid inferences even in situations where those variants only have
weak effects on the exposure. In particular, we exploit the highly
structured nature of genetic correlations in single gene regions to
reduce the dimension of genetic variants using factor analysis. These
genetic factors are then used as instrumental variables to construct
tests for the causal effect of interest. Since these factors may often
be weakly associated with the exposure, size distortions of standard
\textit{t}-tests can be severe. Therefore, we consider two approaches
based on conditional testing. First, we extend results of commonly-used
identification-robust tests to account for the use of estimated factors
as instruments. Secondly, we propose a test which appropriately adjusts
for first-stage screening of genetic factors based on their relevance.
Our empirical results provide genetic evidence to validate cholesterol-lowering
drug targets aimed at preventing coronary heart disease.
\end{abstract}
\begin{center}
\textsc{\small{}Keywords:}{\small{} }\textsl{\small{}cis}{\small{}-Mendelian
randomization, approximate factor models, weak instruments}{\small\par}
\par\end{center}

\newpage{}

\section{Introduction}

Mendelian randomization (MR) is a widely-used method to estimate the
unconfounded effect of an exposure on an outcome by using genetic
variants as instrumental variables. An emerging area of clinical research
concerns MR studies with genetic variants drawn from gene regions
of pharmacological interest. The potential effect of a drug can be
investigated by an MR analysis of a genomic locus (\textsl{cis}-MR)
encoding protein targets of medicines \citep{Walker2017}. As a result,
the \textsl{cis}-MR approach is being increasingly used to provide
valuable evidence which can inform designs of expensive clinical trials
\citep{Gill2021}.

Furthermore, \textsl{cis}-MR approaches that integrate expression
data, with proteins acting as the exposure, are more likely to satisfy
the \textsl{exclusion restriction} required for instrument validity
\citep{Schmidt2019}. The exclusion restriction requires that any
association between instruments and the outcome is only through their
effects on the exposure, which is more plausible when the exposure
is a direct biological product of the instrument.

A starting point for any MR analysis is to choose appropriate instruments.
Since genome-wide association studies (GWASs) have been able to identify
many strong genetic signals for a wide range of traits, the typical
practice in polygenic MR, where variants may be chosen from multiple
gene regions, is to select uncorrelated variants with strong measured
associations with the exposure. 

In contrast, the potential pool of instruments that we can consider
for \textsl{cis}-MR is more limited in two respects. First, the instruments
are typically in highly structured correlation, owing to how genetic
variants in the same region tend to be inherited together. Secondly,
when the exposure of interest is a gene product, genetic associations
are typically measured from much smaller sample sizes than usual GWASs,
which would leave \textsl{cis}-MR analyses more vulnerable to problems
of weak instrument bias \citep*{Andrews2019}. Therefore, for our
\textsl{cis}-MR focus, we will need to make use of \textsl{many weak}
and \textsl{correlated} instruments.

One intuitive option is to filter out variants such that only a smaller
set of uncorrelated (or weakly correlated) instruments remain. However,
for \textsl{cis}-MR analyses that involve only weak genetic signals,
it would seem important to utilise the explanatory power from all
available variants. Another option is to only select variants with
a strong measured association with the exposure. While this might
avoid problems relating to weak instruments, estimation could be more
vulnerable to a \textsl{winner's curse} bias \citep{Goring2002},
resulting in poor inferences if the additional uncertainty from instrument
selection is not accounted for \citep*{Mounier2021}. 

In this paper, we do not propose selecting specific genetic variants
as instruments, but rather genetic factors. We consider a two-stage
approach. In the first-stage, we exploit the highly structured nature
of genetic correlations in single gene regions to reduce the dimension
of genetic variants. Following \citet{Bai2002}'s approximate factor
model, the variation in a large number of genetic variants is assumed
to be explained by a finite number of latent factors. The estimated
genetic factors are particular linear combinations of genetic variants
which aim to capture all systematic variation in the gene region of
interest. In the second stage, these estimated genetic factors are
used as instrumental variables.

In terms of estimation, this two-stage approach is similar to \citet{Burgess2017}'s
inverse-variance weighted principal components (IVW-PC) method. However,
our focus in this paper is not on estimation, but on making valid
inferences. In this regard, a major drawback with the IVW-PC method
is that fails to provide robust inferences under weak instruments.
This is a concern not only due to the potentially smaller sample sizes
involved in \textsl{cis}-MR analyses, but because the first-stage
dimension reduction of genetic variants is based on their mutual correlation,
and not on the strength of their association with the exposure. Thus,
there is no guarantee that the estimated genetic factors would be
strong instruments. 

To provide valid inferences when estimated genetic factors are weak
instruments, we consider two different approaches based on conditional
testing. The first approach generalises popular identification-robust
tests \citep{Moreira2003,Wang} for our setting with estimated genetic
factors as instruments. Similarly to \citet{Bai2010}'s analysis under
strong instruments, the asymptotic null distributions of the identification-robust
test statistics can be established even when the true genetic factors
are not identified. 

One drawback with the identification-robust approaches is that they
are unable to provide point estimates of the causal effect. In situations
where a few instruments are considerably stronger than others, it
is natural to question whether it might be better to discard those
instruments which are almost irrelevant. In our case, if some of the
estimated genetic factors appear to have very weak associations with
the exposure, then we may consider dropping them, and then proceed
with usual point estimation strategies. Therefore, in the second approach,
we propose a test which appropriately adjusts for first-stage screening
of genetic factors based on their relevance. The test controls the
selective type I error: the error rate of a test of the causal effect
given the selection of genetic factors as instruments \citep{Fithian2007}. 

Our empirical motivation concerns a potential drug target of growing
interest. Cholesteryl ester transfer protein (CETP) inhibitors are
a class of drug that increase high-density lipoprotein cholesterol
and decrease low-density lipoprotein cholesterol (LDL-C) concentrations.
A recent \textsl{cis}-MR analysis by \citet{Schmidt2021} suggests
that CETP inhibition may be an effective drug target for reducing
coronary heart disease risk. 

A simulation study based on real \textsl{CETP} gene summary data illustrates
how both factor-based conditional testing approaches offer reliable
inferences under realistic problems faced in practice: small samples,
invalid instruments, and mismeasured instrument correlations. Our
application complements \citet{Schmidt2021}'s findings by providing
robust genetic evidence that a LDL-C lowering effect of CETP inhibitors
is associated with a lower risk of coronary heart disease. 

We use the following notation and abbreviations: $\overset{P}{\to}$
`converges in probability to'; $\overset{D}{\to}$ `converges in distribution
to'; $\overset{a}{\sim}$ 'is asymptotically distributed as'. For
any sequences $a_{n}$ and $b_{n}$, if $a_{n}=O(b_{n})$, then there
exists a positive constant $C$ and a positive integer $N$ such that
for all $n\geq N$, $b_{n}>0$ and $\vert a_{n}\vert\leq Cb_{n}$.
If $a_{n}=o(b_{n})$, then $\vert a_{n}\vert/b_{n}\to0$ as $n\to\infty$.
Also, if $a_{n}=\Theta(b_{n})$, then there exist positive constants
$C_{1}$ and $C_{2}$, $C_{1}\leq C_{2}<\infty$, and a positive integer
$N$ such that $C_{1}b_{n}\leq a_{n}\leq C_{2}b_{n}$ for all $n\geq N$.
Let $(A)_{j}$ denote the $j$-th element of any vector $A$, and
$(B)_{jk}$ denote the $(j,k)$-th element of any matrix $B$. Let
$\Vert.\Vert$ denote the Euclidean norm of a vector. For any positive
integer $A$, $[A]=\{1,\ldots,A\}$. The proofs of the theoretical
results are given in Supplementary Material, and R code to apply our
methods is available at https://github.com/ash-res/con-cis-MR/.

\section{Approximate factor model and summary data}

Let $X$ denote the exposure, $Y$ the outcome, and $Z=(Z_{1},\ldots,Z_{p})^{\prime}$
a vector of $p$ genetic variants which we assume are valid instrumental
variables. For each variant $j\in[p]$, let $\beta_{X_{j}}$ denote
the true marginal association of variant $j$ with the exposure, and
$\beta_{Y_{j}}$ denote the true marginal association of variant $j$
with the outcome. The causal effect of the exposure on the outcome
is denoted $\theta_{0}$, and is described by the linear model 
\begin{equation}
\beta_{Y_{j}}=\theta_{0}\beta_{X_{j}},\,j\in[p]
\end{equation}
Although this specification does not explicitly allow for variants
to have direct effects on the outcome that are not mediated by the
exposure, we will later discuss how this can be relaxed.

\subsection{Two-sample summary data}

We work within the popular two-sample summary data design, where estimated
variant--exposure associations are obtained from a non-overlapping,
but representative, sample from estimated variant--outcome associations.
For each variant, we observe the estimated variant--exposure association
$\hat{\beta}_{X_{j}}$ and a corresponding standard error $\sigma_{X_{j}}$,
from the marginal regression of $X$ on $Z_{j}$, $j\in[p]$. Likewise,
we define $\hat{\beta}_{Y_{j}}$ and $\sigma_{Y_{j}}$ for variant--outcome
associations. These association estimates are often publicly available
from GWASs. For any two variants $j$ and $k$, we also observe the
genetic correlation $\rho_{jk}$, which can be obtained from popular
MR software packages \citep{Hemani2020}. 

Let $\hat{\beta}_{X}=(\hat{\beta}_{X_{1}},\ldots,\hat{\beta}_{X_{p}})^{\prime}$
and $\sigma_{X}=(\sigma_{X_{1}},\ldots,\sigma_{X_{p}})^{\prime}$.
$\hat{\beta}_{Y}$ and $\sigma_{Y}$ are defined analogously. Let
$n_{X}$ denote the size of the sample used to compute variant--exposure
associations, and $n_{Y}$ denote the sample size used to compute
variant--outcome associations. We assume that $n:=n_{X}=cn_{Y}$,
for some positive constant $0<c<\infty$; this ensures that for our
two-sample setting, the sampling uncertainty from one association
study is not negligible to the other. 
\begin{assumption}[\textsc{summary data on genetic associations}]
 For $\beta_{X}=(\beta_{X_{1}},\ldots,\beta_{X_{p}})^{\prime}$ and
$\beta_{Y}=(\beta_{Y_{1}},\ldots,\beta_{Y_{p}})^{\prime}$, $\hat{\beta}_{X}\sim N(\beta_{X},\Sigma_{X})$
and $\hat{\beta}_{Y}\sim N(\beta_{Y},\Sigma_{Y})$, where the $(j,k)$-th
element of $\Sigma_{X}$ is given by $\Sigma_{X_{jk}}=\rho_{jk}\sigma_{X_{j}}\sigma_{X_{k}}$,
the $(j,k)$-th element of $\Sigma_{Y}$ is given by $\Sigma_{Y_{jk}}=\rho_{jk}\sigma_{Y_{j}}\sigma_{Y_{k}}$.
Under the two-sample design, $\hat{\beta}_{X}$ and $\hat{\beta}_{Y}$
are independent. Furthermore, $\Sigma_{X}$ and $\Sigma_{Y}$ are
assumed known, and satisfy $\Sigma_{X}=\Theta(1\big/n)$ and $\Sigma_{Y}=\Theta(1\big/n)$. 
\end{assumption}
We assume that the genetic association estimates are normally distributed
around the true associations, which can be justified by large random
sampling in GWASs. The specific form of the variance-covariance matrices
$\Sigma_{X}$ and $\Sigma_{Y}$ relies on an assumption that the true
genetic associations are quite weak. The assumption that the standard
errors $\sigma_{X_{j}}$ and $\sigma_{Y_{j}}$, $j\in[p]$ are known,
and decrease at the usual parametric rate, is common in practice \citep[Assumption 1, p.3]{Zhao2018},
and is thought not to be too restrictive \citep[Theorem 4.1, p.9]{Ye2020}.
The remaining components in $\Sigma_{X}$ and $\Sigma_{Y}$ are the
pairwise genetic correlations $\rho_{jk}$, $j,k\in[p]$. We assume
these are known, but our results can be extended to allow for consistently
estimated genetic correlations. 

\subsection{Approximate factor model}

Our asymptotic framework considers the setting where $p\to\infty$,
since we aim to incorporate information from many genetic variants.
For our \textsl{cis}-MR focus, we would expect a large number of genetic
variants to be in highly structured correlation. Therefore, we assume
genetic variants in the region of interest follow an approximate factor
model structure \citep{Bai2002}. For the vector of genetic variants
$Z=(Z_{1},...,Z_{p})^{\prime}$, we have 
\begin{equation}
Z=\Lambda f+e,
\end{equation}
where $\Lambda=(\lambda_{1},...,\lambda_{p})^{\prime}$ is an unobserved
$p\times r$ matrix of factor loadings, $f$ is a $r$-vector of unobserved
factors, and $e=(e_{1},...,e_{p})^{\prime}$ is a $p$-vector of idiosyncratic
errors. The component $\lambda_{j}^{\prime}f$ describes the systematic
variation in any $j$-th genetic variant $Z_{j}$. Although $p\to\infty$,
$r$ is considered to be finite; the systematic variation of $p$
variants can be explained by a much smaller set of $r$ latent factors.
Thus, instead of using $p$ genetic variants as instruments, we will
aim to use the information of these $r$ latent factors to construct
instruments. 
\begin{assumption}[\textsc{approximate factor model}]
\textsl{\emph{ }}(i) the unobserved factors and idiosyncratic errors
are identically and independently distributed across individuals;
(ii) the idiosyncratic errors may have some dependence across variants;
(iii) the idiosyncratic errors may have some correlation with the
sampling errors of genetic association estimates; (iv) the factors
satisfy $\mathbb{E}[\Vert f\Vert^{4}]=O(1)$ and $\Sigma_{F}=\mathbb{E}[ff^{\prime}]$
is an $r\times r$ positive definite matrix; (iv) $\Vert\lambda_{j}\Vert\leq C_{\lambda}$
for some constant $C_{\lambda}>0$, and $p^{-1}\Lambda^{\prime}\Lambda\to\Sigma_{\Lambda}$,
where $\Sigma_{\Lambda}$ is a positive definite, non-random matrix,
as $p\to\infty$. 
\end{assumption}
Assumption 2 implies Assumptions A-F from \citet[pp.141-4]{Bai2003};
the assumptions imply $r$ strong factors exist, and that, as $p\to\infty$,
there is a significant difference between the $r$-th and $(r+1)$-th
eigenvalues of the genetic correlation matrix $\rho$. Compared with
classical factor models, the assumptions maintained in an approximate
factor model are weak enough to prevent separate identification of
factors and factor loadings, however both can be estimated up to a
$r\times r$ rotation matrix. This should involve no loss of information
since in terms of retaining the same explanatory power, we only require
that the estimated factors span the same space as the true factors
\citep{Bai2002}. For details on the dependence allowed between idiosyncratic
errors across variants, see \citet[Assumption A(c), p.1581]{Bai2010}.
Further details on the correlations permitted between the idiosyncratic
errors in $(2)$ and the sampling errors of genetic association estimates
are provided in Supplementary Material.

\subsection{Weak genetic associations}

The focus of this paper is to develop methods for \textsl{cis}-MR
inference which are robust to weak genetic signals in the gene region
of interest. We believe such methods are particularly important for
the potential of \textsl{cis}-MR to identify novel therapeutic targets,
since strong genetically-proxied effects of the exposure may not yet
be established, and studies may be under-powered due to small samples.
Our asymptotic analysis relies on the assumption that many variants
have weak genetic associations with the exposure; see for example,
\citet{Zhao2018}. 
\begin{assumption}[\textsc{many weak genetic associations}]
 $p\big/n=\Theta(1)$ and $\Vert\beta_{X}\Vert=O(1)$ as $n,p\to\infty$. 
\end{assumption}
Assumption 3 implies the average explanatory power of any individual
variant is decreasing with the total number of variants. Since we
do not directly use individual variants as instruments, this does
not necessarily imply we face a weak instruments problem. We will
take $r$ linear combinations of all variants to use as instruments;
as discussed by \citet{Bai2010}, it is possible for these linear
combinations to be strong instruments even if the explanatory power
of individual variants is limited. 

However, for our focus, we deem this to be quite unrealistic: our
dimension reduction of genetic variants will be based on their correlation
structure, not their association with the exposure. Hence, we could
end up using instruments that are able to summarise nearly all genetic
variation in a gene region, but they are still weakly associated with
the exposure. For this reason, we should focus on inferential methods
that are robust to weak instruments. 

\subsection{Example: the effect of statins on coronary heart disease}

There is an well-established association between high low-density
lipoprotein cholesterol (LDL-C) levels and greater risk of coronary
heart disease (CHD); see, for example, \citet{Mihaylova2012}. Statins
are 3-Hydroxy-3-Methylglutaryl-CoA Reductase (HMGCR) inhibitors prescribed
to lower LDL-C levels. A \textsl{cis}-MR analysis using genetic variants
from the \textsl{HMGCR} gene region provides a way to study the genetically-proxied
effect of statin intake on CHD \citep{Ference2016}. 

The left panel of Figure 1 shows how genetic variants in \textsl{HMGCR}
are in strong correlation. Nearly 98 percent of the total variation
of 2883 genetic variants can be summarised by 14 principal components.
With this in mind, we might consider 14 estimated factors as instrumental
variables. Since the construction of factor-based instruments does
not use the information of variant--exposure associations, the estimated
genetic factors may be weak instruments. 
\begin{center}
\includegraphics[width=16cm]{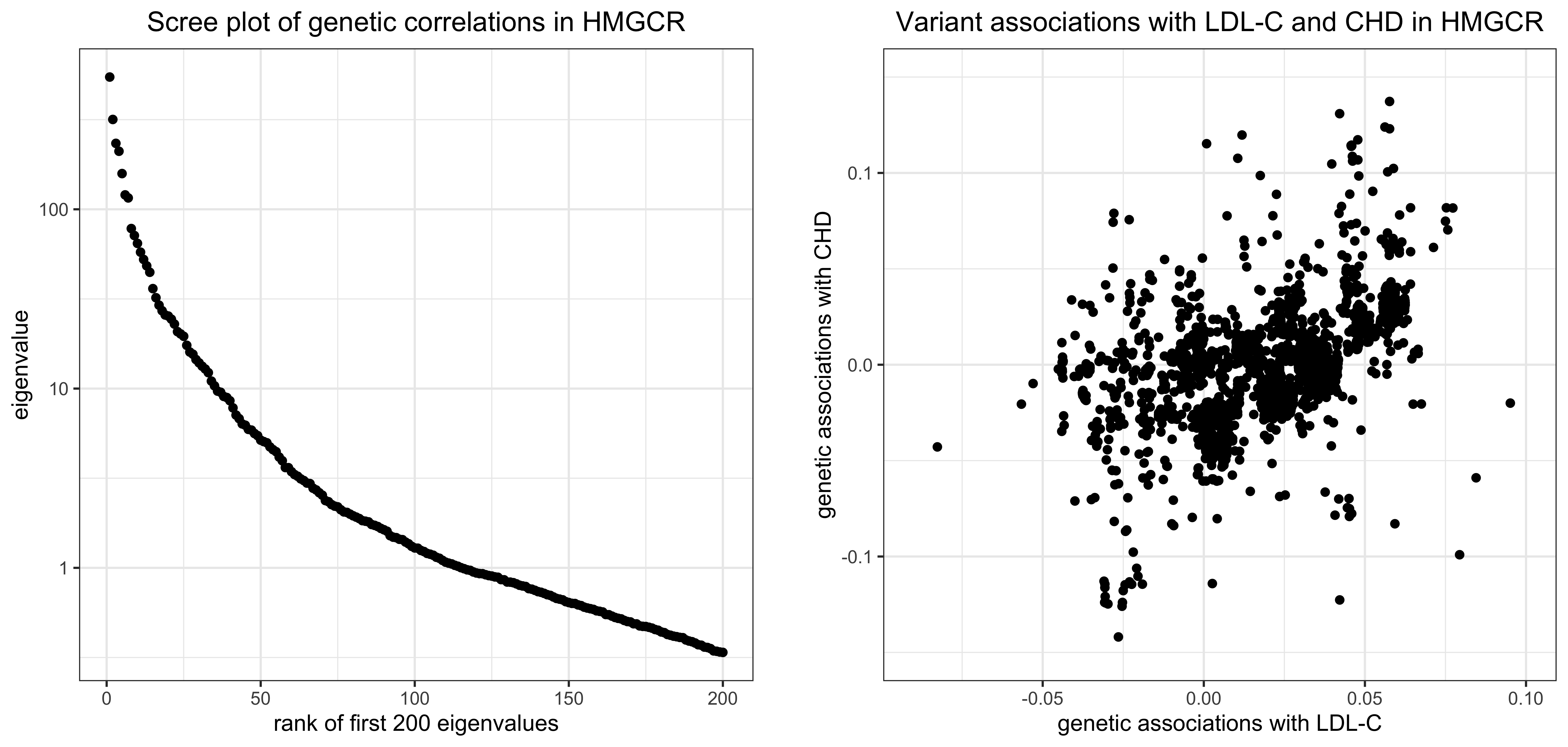}\\
{\footnotesize{}Figure 1. Scree plot of genetic associations (left),
and genetic associations of LDL-C and CHD (right) in the }\textsl{\footnotesize{}HMGCR}{\footnotesize{}
gene region.}{\footnotesize\par}
\par\end{center}

We consider two options to make valid inferences in this setting.
The first option is to use all 14 genetic factors as instruments,
and conduct identification-robust tests which are designed to control
type I error rates under weak instruments. This approach does not
return a point estimate for the causal effect, but provides valid
inferences regardless of instrument strength. The second option is
to discard any genetic factors with very weak measured associations
with the exposure, and to proceed with usual point estimation strategies
assuming that the retained genetic factors are strong instruments.
To make honest inferences with this second option, we can conduct
conditional tests which account for first-stage instrument selection. 

\section{Conditional inference in \textsl{cis}-MR with genetic factors}

\subsection{Estimating the factor loadings}

We start by estimating the $p\times r$ matrix of factor loadings
$\Lambda$ using the variant correlation matrix. Let $\rho$ denote
the variant correlation matrix, so that its $(j,k)$-th element is
given by $\rho_{jk}$. For a given number of factors $r$, let $\bar{\Lambda}$
denote a $p\times r$ matrix with its columns given by the eigenvectors
corresponding to the largest $r$ eigenvalues of $\rho$ multiplied
by $\sqrt{p}$. Then, the estimated (re-scaled) factor loadings are
given by $\hat{\Lambda}=\bar{\Lambda}(p^{-1}\bar{\Lambda}^{\prime}\bar{\Lambda})^{-\frac{1}{2}}$,
so that $p^{-1}\hat{\Lambda}^{\prime}\hat{\Lambda}=I_{r}$. 

For our analysis, the number of factors $r$ is assumed to be known,
for example, by inspecting the scree plot of the variant correlation
matrix. In practice, one could also use data-driven methods to determine
the number of factors \citep{Bai2002,Onatski2010}.

\subsection{Point estimation under strong factor associations}

Under the linear model $\beta_{Y}=\beta_{X}\theta_{0}$, we can use
our variant--exposure and variant--outcome associations to construct
a vector of estimating equations, $\hat{\beta}_{Y}-\hat{\beta}_{X}\theta=0$.
Here, there are $p$ estimating equations for $1$ unknown. 

Given our estimated factor loadings, we can effectively reduce the
degree of over-identification. Let $\hat{g}(\theta)=\hat{\Lambda}^{\prime}(\hat{\beta}_{Y}-\hat{\beta}_{X}\theta)$,
so that $\hat{g}(\theta)=0$ provides $r$ estimating equations for
$\theta_{0}$. In other words, $\hat{g}(\theta)=0$ are the estimating
equations implied by using the linear combination of variants $\hat{\Lambda}^{\prime}Z$
as instruments. For brevity, we will refer to $\hat{\Lambda}^{\prime}Z$
as the \textsl{estimated factors}. 

A consistent estimator of the variance-covariance matrix of $\hat{g}(\theta)$
is given by $\hat{\Omega}(\theta)=\hat{\Lambda}^{\prime}(\Sigma_{Y}+\theta^{2}\Sigma_{X})\hat{\Lambda}$.
Then, a limited information maximum likelihood (LIML; \citealp{Anderson1949})
estimator is given by
\[
\hat{\theta}_{F}=\arg\min_{\theta}\hat{g}(\theta)^{\prime}\hat{\Omega}(\theta)^{-1}\hat{g}(\theta).
\]
We call $\hat{\theta}_{F}$ the F-LIML estimator; the LIML estimator
which uses the entire vector of estimated factors as instruments.
\begin{thm}[\textsc{estimation with strong instruments}]
 Under Assumptions 1-3 and Equation (1), if $\Lambda^{\prime}\beta_{X}=\Theta(p^{\frac{1}{2}})$,
then $\hat{V}^{-\frac{1}{2}}(\hat{\theta}_{F}-\theta_{0})\overset{D}{\to}N(0,1)$
as $n,p\to\infty$, where $\hat{V}=(\hat{G}^{\prime}\hat{\Omega}^{-1}\hat{G})^{-1}$,
$\hat{G}=-\hat{\Lambda}^{\prime}\hat{\beta}_{X}$, and $\hat{\Omega}=\hat{\Omega}(\hat{\theta})$. 
\end{thm}
Under the condition that all estimated factors are strong instruments,
we can directly use Theorem 1 to construct asymptotic confidence intervals
and tests for the causal effect $\theta_{0}$. 

\subsection{Identification-robust tests under weak factor associations}

Standard $t$-tests based on Theorem 1 will not be valid when the
estimated factors are weak instruments. This is because under weak
instrument asymptotics the distribution of $t$-tests will depend
on a measure of instrument strength \citep{Stock2002}. Instead, identification-robust
tests offer a way to make valid inferences in this setting. The basic
idea behind this approach is to construct pivotal test statistics
\textsl{conditional} on a sufficient statistic for instrument strength.
Then, since the conditional distributions of these test statistics
do not depend on instrument strength under the null hypothesis, the
size of the tests can be controlled under weak instruments.

We can follow previous works by constructing these test statistics
as a function of two asymptotically mutually independent statistics
$\bar{S}$ and $\bar{T}$, where $\bar{S}$ carries the information
of the estimated factors being valid instruments, and where $\bar{T}$
incorporates information on the strength of these instruments. Specifically,
under the null hypothesis ${\cal H}_{0}:\theta=\theta_{0}$, let $\bar{S}=\hat{\Omega}(\theta_{0})^{-\frac{1}{2}}\hat{g}(\theta_{0})$,
and $\bar{T}=(\hat{\Omega}_{X}-\hat{\Delta}_{G}\hat{\Omega}(\theta_{0})^{-1}\hat{\Delta}_{G})^{-\frac{1}{2}}(\hat{G}-\hat{\Delta}_{G}\hat{\Omega}(\theta_{0})^{-1}\hat{g}(\theta_{0}))$,
where $\hat{\Omega}_{X}=\hat{\Lambda}^{\prime}\Sigma_{X}\hat{\Lambda}$,
$\hat{\Delta}_{G}=\hat{\Omega}_{X}\theta_{0}$, and where $\hat{\Omega}(\theta_{0})$
is defined in Section 3.2. 

Using $\bar{S}$ and $\bar{T}$, we can construct three commonly-used
identification-robust test statistics which will be asymptotically
pivotal conditional on ${\cal Z}_{T}\sim N((\Omega_{X}-\Delta_{G}\Omega^{-1}\Delta{}_{G})^{-\frac{1}{2}}G,I_{r})$,
where $\Omega_{X}=H^{-1}\Lambda^{\prime}\Sigma_{X}\Lambda(H^{-1})^{\prime}$,
$\Delta_{G}=\Omega_{X}\theta_{0}$, $\Omega=H^{-1}\Lambda^{\prime}(\Sigma_{Y}+\theta_{0}^{2}\Sigma_{X})\Lambda(H^{-1})^{\prime}$,
$G=-H^{-1}\Lambda^{\prime}\beta_{X}$, and $H$ is a rotation matrix.
Let $\bar{Q}_{S}=\bar{S}^{\prime}\bar{S}$, $\bar{Q}_{ST}=\bar{S}^{\prime}\bar{T}$,
and $\bar{Q}_{T}=\bar{T}^{\prime}\bar{T}$. Then, the \citet{Anderson1949}
statistic with estimated factors is given by $\text{F-AR}=\bar{Q}_{S}$,
\citet{Kleibergen2005}'s Lagrange multiplier statistic with estimated
factors is given by $\text{F-LM}=\bar{Q}_{ST}^{2}/\bar{Q}_{T}$, and
\citet{Moreira2003}'s conditional likelihood ratio statistic with
estimated factors is given by $\text{F-CLR}=\big(\bar{Q}_{S}-\bar{Q}_{T}+\big[(\bar{Q}_{S}+\bar{Q}_{T})^{2}-4(\bar{Q}_{S}\bar{Q}_{T}-\bar{Q}_{ST}^{2})\big]^{\frac{1}{2}}\big)\big/2$. 
\begin{thm}[\textsc{identification-robust test statistics}]
 Suppose that $\Lambda^{\prime}\beta_{X}=\Theta(1)$ and $\beta_{X_{k}}=\Theta(p^{-\frac{1}{2}})$,
$k\in[p]$. Under Assumptions 1-3, Equation (1) and ${\cal H}_{0}:\theta=\theta_{0}$,
conditional on ${\cal Z}_{T}$, \textbf{\emph{(i)}} $\text{F-AR}\overset{D}{\to}\chi_{r}^{2}$;
\textbf{\emph{(ii)}} $\text{F-LM}\overset{D}{\to}\chi_{1}^{2}$; and
\textbf{\emph{(iii)}} $\text{F-CLR}\overset{D}{\to}\big(\chi_{1}^{2}+\chi_{r-1}^{2}-{\cal Z}_{T}^{\prime}{\cal Z}_{T}+[(\chi_{1}^{2}+\chi_{r-1}^{2}-{\cal Z}_{T}^{\prime}{\cal Z}_{T})^{2}+4\chi_{1}^{2}{\cal Z}_{T}^{\prime}{\cal Z}_{T}]^{\frac{1}{2}}\big)\big/2$
as $n,p\to\infty$, where $\chi_{1}^{2}$ and $\chi_{r-1}^{2}$ denote
independent chi-square random variables. 
\end{thm}
Since the F-AR statistic is not a function of $\bar{T}$, it does
not incorporate the identifying power of instruments. As a result,
when the model is over-identified ($r>1$), the F-AR test may have
relatively poor power properties compared with the F-LM and F-CLR
tests \citep*{Andrews2019}. Of the three methods, CLR-based tests
are widely regarded as the most powerful, due to simulation evidence
\citep*{Andrews2007a} and favourable theoretical properties \citep*{Andrews2006}. 

\subsection{Conditional tests that adjust for factor selection}

While identification-robust tests are designed to control type I error
rates for any level of instrument strength, in a sparse effects setting
where a few estimated factors are strong instruments and most other
estimated factors are very weak instruments, it would be tempting
to proceed with F-LIML point estimation after removing very weak instruments.
Without any instrument selection, we could use the entire $r$-vector
of estimated factors as instruments, as in Sections 3.2 and 3.3. In
contrast, here we wish to filter out certain elements if they are
demonstrably weak instruments. 

To this end, we construct pre-tests to identify a subset of estimated
factors that pass of threshold of relevance; only this subset are
then used as instruments. By \citet{Bai2003}, $\hat{\Lambda}$ estimates
$\Lambda H^{\prime-1}$ where $H$ is a rotation matrix. Hence, the
estimated factor associations $\hat{\Lambda}^{\prime}\hat{\beta}_{X}$
actually estimate $H^{-1}\Lambda^{\prime}\beta_{X}$, and not $\Lambda^{\prime}\beta_{X}$
as we may intuitively expect. Therefore, for each $j\in[r]$, we will
test the null hypothesis ${\cal H}_{0j}:(G)_{j}=0$ against the alternative
${\cal H}_{1j}:(G)_{j}\not=0$, where $G=-H^{-1}\Lambda^{\prime}\beta_{X}$. 

Simple \textsl{t}-tests are used to screen for relevant estimated
factors, using the asymptotic approximation $\hat{\Omega}_{X}^{-\frac{1}{2}}(\hat{G}-G)\overset{a}{\sim}N(0,I_{r})$.
In particular, to conduct a two-sided asymptotic $\delta$-level test
for the significance of each estimated factor $j\in[r]$, we compare
the test statistic $\vert{\cal \hat{T}}_{j}\vert$, where ${\cal \hat{T}}_{j}=(\hat{\Omega}_{X})_{jj}^{-\frac{1}{2}}(\hat{G})_{j}$,
against the critical value $c_{\delta}:=\Phi(1-\delta/2)$, where
$\Phi(.)$ is the standard normal cumulative distribution function.
For each $j\in[r]$, if $\vert{\cal \hat{T}}_{j}\vert>c_{\delta}$,
then we have evidence to reject ${\cal H}_{0j}$, and thus we include
the estimated factor $j$ as an instrument. Any estimated factors
such that $\vert{\cal \hat{T}}_{j}\vert\leq c_{\delta}$ are deemed
to be weak instruments, and are thus discarded. 

Let ${\cal S}$ denote the selection event according to these pre-tests.
For example, if $r=3$ and only the first and third estimated factors
pass the pre-test of relevance, then ${\cal S}=\{\vert{\cal \hat{T}}_{1}\vert>c_{\delta},\vert{\cal \hat{T}}_{2}\vert\leq c_{\delta},\vert{\cal \hat{T}}_{3}\vert>c_{\delta}\}$.
Only using the subset of estimated factors that have passed the pre-test
of relevance, we will test the null hypothesis ${\cal H}_{0}:\theta=\theta_{0}$
against the general alternative ${\cal H}_{1}:\theta\not=\theta_{0}$.
To appropriately account for pre-testing of relevant estimated factors,
we seek to construct a conditional test which controls the selective
type I error \citep{Fithian2007}. That is, for an $\alpha$-level
test, $\mathbb{P}(\text{reject}\,{\cal H}_{0}\vert{\cal S})\leq\alpha$
under ${\cal H}_{0}$; i.e. we control the error rate of the test
at $\alpha$ \textsl{given} the selection event ${\cal S}$. 

Suppose $r^{\star}$ is the number of selected estimated factors,
and let ${\cal R}$ denote the indices of the selected set of estimated
factors, so that the selection event is ${\cal S}=\{\vert{\cal \hat{T}}_{j}\vert>c_{\delta},j\in{\cal R}\}\cap\{\vert{\cal \hat{T}}_{k}\vert\leq c_{\delta},k\in[r]\backslash{\cal R}\}$.
We also let $\Gamma_{{\cal S}}$ denote an $r\times r^{\star}$ selection
matrix which is constructed such that $\hat{\Lambda}\Gamma_{{\cal S}}$
is the $p\times r^{\star}$ matrix where its $r^{\star}$ columns
are the columns of $\hat{\Lambda}$ that correspond to the selected
factors ${\cal R}$. We call the resulting LIML estimator which uses
only the selected factors `Selected LIML' (S-LIML), which is denoted
$\hat{\theta}_{{\cal S}}$. 

To make honest inferences, we will need to consider how the distribution
of $\hat{\theta}_{{\cal S}}$ is impacted by the uncertainty of the
pre-test results ${\cal S}$. The asymptotic conditional distribution
of $\hat{\theta}_{{\cal S}}$ given ${\cal S}$ depends on an $r$-dimensional
nuisance parameter $D^{-\frac{1}{2}}G$, where $D$ is the diagonal
matrix with its $(j,j)$-th element given by $(\Omega_{X})_{jj}$.
Thus, along with the selection event ${\cal S}$, we will condition
on a sufficient statistic for the nuisance parameter, which will cause
it to drop from the conditional distribution of $\hat{\theta}_{{\cal S}}$
\citep[see, for example,][]{Sampson2005}. 

We define some additional quantities in order to introduce a sufficient
statistic for the nuisance parameter. The conditional covariance of
the vector of pre-test statistics $({\cal \hat{T}}_{1},\ldots,{\cal \hat{T}}_{r})^{\prime}$
and the estimate $\hat{\theta}_{{\cal S}}$ is given by $C_{G}=-D^{-\frac{1}{2}}\Omega_{X}\Gamma_{{\cal S}}\Omega_{{\cal S}}^{-1}G_{{\cal S}}V_{{\cal S}}\theta_{0}$,
where $V_{{\cal S}}=(G_{{\cal S}}^{\prime}\Omega_{{\cal S}}^{-1}G_{{\cal S}})^{-1}$,
$G_{{\cal S}}=\Gamma_{{\cal S}}^{\prime}G$, and $\Omega_{{\cal S}}=\Gamma_{{\cal S}}^{\prime}\Omega\Gamma_{{\cal S}}$.
Then, a sufficient statistic for $D^{-\frac{1}{2}}G$ is given by
$U=D^{-\frac{1}{2}}\hat{G}-C_{G}V_{{\cal S}}^{-1}(\hat{\theta}_{S}-\theta_{0})$. 
\begin{thm}[\textsc{conditional distribution of S-LIML estimators}]
 Under Assumptions 1-3, Equation (1) and ${\cal H}_{0}:\theta=\theta_{0}$,
$G_{{\cal S}}=\Theta(p^{\frac{1}{2}})$ and $\beta_{X_{k}}=\Theta(p^{-\frac{1}{2}})$,
$k\in[p]$, the asymptotic conditional distribution of $\hat{\theta}_{{\cal S}}$
given $U=u$ and ${\cal S}$ is approximately 
\begin{equation}
\mathbb{P}(\hat{\theta}_{{\cal S}}\leq w+\theta_{0}\vert{\cal S},U=u)\overset{a}{\sim}\frac{\mathbb{P}\Big(\{V_{{\cal S}}^{\frac{1}{2}}{\cal K}\leq w\}\bigcap_{j\in{\cal R}}\{\vert(\bar{u})_{j}\vert>c_{\delta}\}\bigcap_{k\in[r]\backslash{\cal R}}\{\vert(\bar{u})_{k}\vert\leq c_{\delta}\}\Big)}{\mathbb{P}\Big(\bigcap_{j\in{\cal R}}\{\vert(\bar{u})_{j}\vert>c_{\delta}\}\bigcap_{k\in[r]\backslash{\cal R}}\{\vert(\bar{u})_{k}\vert\leq c_{\delta}\}\Big)},
\end{equation}
as $n,p\to\infty$, where $\bar{u}=u+C_{G}V_{{\cal S}}^{-\frac{1}{2}}{\cal K}$,
and ${\cal K}\sim N(0,1)$. 
\end{thm}
Intuitively, this conditional distribution of $\hat{\theta}_{{\cal S}}$
reveals what the likely values of $\hat{\theta}_{{\cal S}}$ \textsl{should}
be under ${\cal H}_{0}:\theta=\theta_{0}$, given the results of the
pre-tests ${\cal S}$ and observed value $U=u$. If the S-LIML estimate
$\hat{\theta}_{{\cal S}}$ does not lie in a suitable likely region,
then we interpret this as evidence against ${\cal H}_{0}$. 

Let $\hat{D}$ denote an $r\times r$ diagonal matrix with its $(j,j)$-th
element given by $(\hat{\Omega}_{X})_{jj}$. We can construct consistent
estimators $\hat{C}_{G}=-\hat{D}^{-\frac{1}{2}}\hat{\Omega}_{X}\Gamma_{{\cal S}}\hat{\Omega}_{{\cal S}}^{-1}\hat{G}_{{\cal S}}\hat{V}_{{\cal S}}\hat{\theta}_{{\cal S}}$
of $C_{G}$ and $\hat{V}_{{\cal S}}=(\hat{G}_{{\cal S}}^{\prime}\hat{\Omega}_{{\cal S}}^{-1}\hat{G}_{{\cal S}})^{-1}$
of $V_{{\cal S}}$, where $\hat{G}_{{\cal S}}=\Gamma_{{\cal S}}^{\prime}\hat{G}$,
and $\hat{\Omega}_{{\cal S}}=\Gamma_{{\cal S}}^{\prime}\hat{\Omega}(\hat{\theta}_{{\cal S}})\Gamma_{{\cal S}}$.
By conditioning on $u=\hat{D}^{-\frac{1}{2}}\hat{G}-\hat{C}_{G}\hat{V}_{{\cal S}}^{-1}(\hat{\theta}_{{\cal S}}-\theta_{0})$,
we can conduct an approximate $\alpha$-level test for ${\cal H}_{0}:\theta=\theta_{0}$
by using the sample analogue of the right-hand side of Equation $(3)$,
taking repeated draws of ${\cal K}\sim N(0,1)$, and computing $\alpha/2$
and $(1-\alpha/2)$-level quantiles of the approximated $\mathbb{P}(\hat{\theta}_{{\cal S}}\leq w+\theta_{0}\vert{\cal S},U=u)$
distribution under ${\cal H}_{0}$. If the S-LIML estimate $\hat{\theta}_{{\cal S}}$
does not lie within those quantiles, then we reject the null hypothesis
${\cal H}_{0}$. 

\section{Simulation study}

This section presents the performance of the identification-robust
and conditional test statistics in a simulation study based on real
genetic data. The simulation design aims to explore the robustness
of our empirical results in Section 5, where we investigate cholesteryl
ester transfer protein (CETP) inhibitors as a potential drug target
for coronary heart disease (CHD). CETP are a class of drug which increase
high-density lipoprotein cholesterol and decrease low-density lipoprotein
(LDL-C). Therefore, our exposure $X$ is LDL-C, our outcome $Y$ is
CHD, and our potential pool of instruments $Z$ are a set of 196 genetic
variants from taken a neighborhood of the \textsl{CETP} gene which
are associated with LDL-C at p-value less than $5\times10^{-4}$. 

The genetic associations with LDL-C and CHD are generated from two
independent GWASs using the normality assumptions in Assumption 1.
The true variant--exposure associations $\beta_{X}$ are set as the
measured variant associations with LDL-C, and the true variant--outcome
associations $\beta_{Y}$ are set as the measured variant associations
with CHD. The covariance matrix for $\hat{\beta}_{X}$ is formed as
$\Sigma_{X}=\rho\circ\sigma_{X}\sigma_{X}^{\prime}$, where $\rho$
is an estimated genetic correlation matrix of the 196 variants, and
$\sigma_{X}$ is the vector of standard errors associated with the
measured variant associations with LDL-C. The covariance matrix $\Sigma_{Y}$
for $\hat{\beta}_{Y}$ is formed analogously. 

Across the range of scenarios we considered, the CLR-based tests were
generally the most reliable and powerful. Therefore, our discussion
here focuses on the results of the CLR tests, and we will omit presenting
the results of the AR and LM-based tests for clarity. 

Our proposed methods are based on dimension reduction of all variants
rather than selecting specific variants. Instead of using estimated
factors as instruments, we might wonder if it is better to simply
omit highly correlated variants and conduct identification-robust
tests with a smaller number of moderately correlated variants as instruments.
This is the approach studied by \citet{Wang}. Thus, a natural source
of comparison with our factor-based methods are CLR tests which filter
out variants if they are correlated with an already included variant
at some pre-specified threshold. In MR terminology, this is called
\textsl{pruning}. The tests CLR-20, CLR-40, CLR-60, and CLR-80, will
denote CLR tests computed with pruned variants such that their correlations
are bounded at $R^{2}$ thresholds of $0.2$, $0.4$, $0.6$, and
$0.8$, respectively. We also compute CLR-01, which is an CLR test
using a set of mutually almost uncorrelated variants ($R^{2}\leq0.01)$,
after choosing the most strongly associated variant with LDL-C. 

To conduct our S-LIML and F-CLR tests, we need to decide on an appropriate
number of factors $r$. For our simulations, we chose $r=8$ since
there was a noticeable gap between the $8$-th and $9$-th eigenvalues
of the variant correlation matrix $\rho$.

Our simulation study focuses on studying the performance of tests
under three practical problems of interest in \textsl{cis}-MR analyses:
small sample sizes, invalid instruments, and mismeasured instrument
correlations. 

\subsection{Smaller sample sizes}

Since proteins are the drug target of most medicines, \textsl{cis}-MR
analyses often use proteins as the exposure of interest. Genetic associations
with protein or gene expression are typically measured with smaller
sample sizes than usual GWASs. In practice, this can result in a weak
instruments problem since the reported standard errors will be larger.
To mimic a small sample size problem, instead of taking the reported
standard errors of variant--LDL-C and variant--CHD associations
at face value, we divide the standard errors by a fixed constant $0<\eta\leq1$.
Hence, we generate two-sample summary data as $\hat{\beta}_{X}\sim N(\bar{\beta}_{X},\eta^{-1}\Sigma_{X})$
and $\hat{\beta}_{Y}\sim N(\theta_{0}\bar{\beta}_{X},\eta^{-1}\Sigma_{Y})$,
where $\bar{\beta}_{X}$ are the reported variant--LDL-C associations.
A smaller value of $\eta$ implies a smaller sample size. 
\begin{center}
\includegraphics[width=16.5cm]{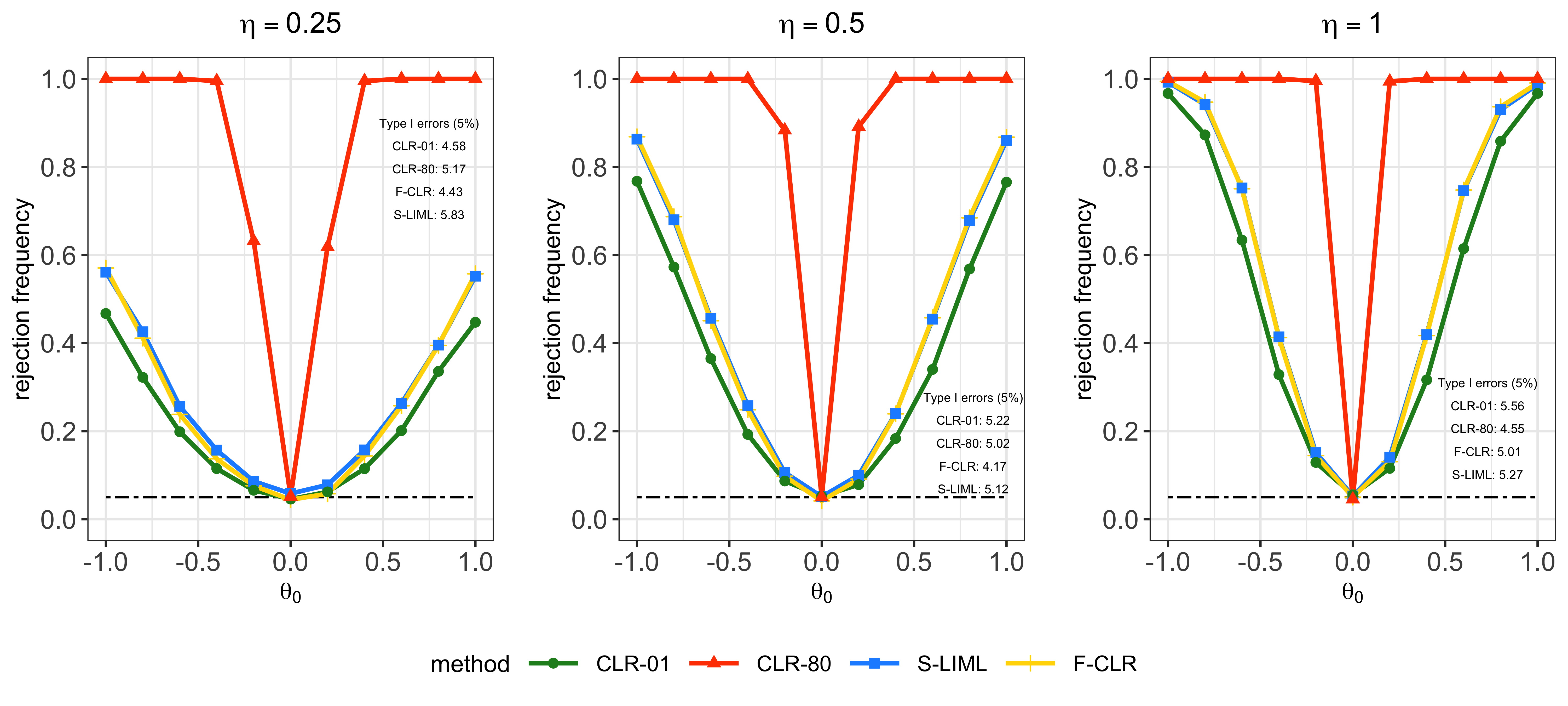}{\footnotesize{}}\\
{\footnotesize{}Figure 2. Power with small samples when testing the
null hypothesis ${\cal H}_{0}:\theta_{0}=0$. }{\footnotesize\par}
\par\end{center}

Figure 2 shows that when all variants are valid instruments, using
very correlated individual variants as instruments perform well; the
CLR test with a pruning threshold of $R^{2}=0.8$ (CLR-80) performed
best. Using estimated factors as instruments provides reliable inference
for both F-CLR and S-LIML approaches, with both tests performing better
than the CLR-01 test which only uses uncorrelated variants as instruments,
although S-LIML is over-sized in very small samples ($\eta=0.25$).
For this, and future simulation designs, the S-LIML test screened
the estimated factors using $\delta=0.1$, $\delta=0.05$, $\delta=0.01$
level pre-tests for the $\eta=0.25$, $\eta=0.5$, and $\eta=1$ scenarios,
respectively. With these thresholds most of the 8 estimated factors
were retained as instruments, but usually not all of them. All $8$
factors were selected in $34.4\%$ of cases under $\eta=1$, $17.9\%$
of cases under $\eta=0.5$, and just $5.6\%$ of cases under $\eta=0.25$. 

The inflated type I error rates of the S-LIML approach under very
small samples is perhaps not a surprise; while the CLR-based approaches
are robust to weak instruments, LIML-based point estimators are not.
Even if we appropriately screen out very weak instruments, if the
strongest instrument is still quite weak, then the usual asymptotic
approximations used to carry out inferences can be poor \citep{Stock2002}. 

\subsection{Invalid instruments}

While the linear model $(1)$ is thought to provide a reasonable approximation
to practice, we may be concerned that proportionality of genetic associations
$\beta_{Y_{j}}=\theta_{0}\beta_{X_{j}}$ may not hold exactly over
all variants $j\in[p]$. Fortunately, the factor model approach may
provide some robustness to the inclusion of invalid instruments. For
example, under \citet{Bai2010}'s analysis, a finite number of variants
would be permitted to have direct effects on the outcome as long as
the total number of variants $p$ grows sufficiently faster than $n$.
\begin{center}
\includegraphics[width=16.5cm]{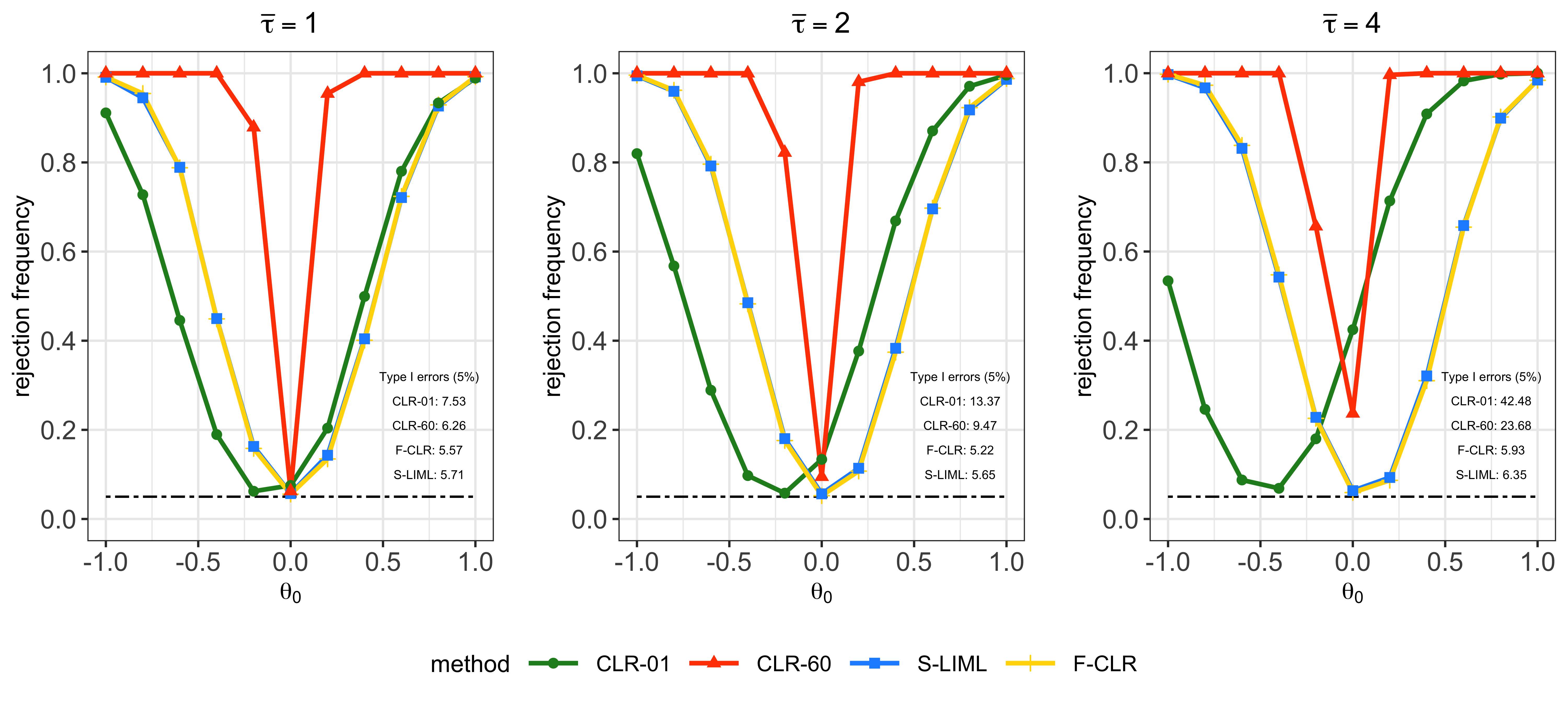}{\footnotesize{}}\\
{\footnotesize{}Figure 3. Power with invalid instruments when testing
the null hypothesis ${\cal H}_{0}:\theta_{0}=0$. }{\footnotesize\par}
\par\end{center}

Here we study finite-sample behaviour under the model $\beta_{Y_{j}}=\theta_{0}\beta_{X_{j}}+\tau_{j}$,
where the direct effects are generated as $\tau_{j}\sim U[-0.005\bar{\tau},0.005\bar{\tau}]$,
$\tau>0$, $j\in[p]$. This is similar to an assumption of balanced
pleiotropy often maintained in polygenic MR \citep{Hemani2018}. When
the direct effects $\tau_{j}$ are random around zero, their impact
is to inflate the variance of the resulting estimate. Therefore, when
the distribution of the random effects $\tau_{j}$ is known, valid
inferences can be obtained by estimating a variance correction term
to account for the extra heterogeneity \citep{Zhao2018,Burgess2020}.
In contrast, here we set the direct effects $\tau_{j}$ to be fixed,
so that they directly impact the bias of the resulting estimates,
with no de-biasing adjustment possible without imposing further restrictions. 

Figure 3 highlights the robust performance of the F-CLR and S-LIML
tests under invalid instruments. F-CLR was the best performing test,
with only small size distortions. S-LIML was slightly more over-sized
with seemingly no power advantage over F-CLR for our experiments based
on \textsl{CETP} gene data. The pruning approaches were generally
much less robust to invalid instruments, showing substantially inflated
type I error rates. Intuitively, we would expect that more liberal
pruning thresholds might perform better, as they allow for the direct
effects to average out. Of the 5 pruning thresholds we considered,
a $R^{2}=0.6$ threshold performed the best, with CLR-60 making use
of 40 out of the 196 variants available. However, even CLR-60 was
heavily over-sized when the direct effects were large ($\bar{\tau}\geq2$). 

\subsection{Mismeasured variant correlations}

Our final simulation design investigates robustness to a very common
problem in \textsl{cis}-MR analyses. It is often the case that the
variant correlation matrix $\rho$ is not provided alongside summary
genetic association data. In such situations, if researchers want
to make use of correlated variants, they would need to obtain estimates
of the variant correlation matrix from a different set of subjects.
This was the case for our \textsl{CETP} gene analysis, where we obtained
a variant correlation matrix $\bar{\rho}$ from a reference panel
\citep[1000 Genomes Project,][]{Auton2015} using the MR-Base platform
\citep{Hemani2020}. 

Discrepencies between the variant correlation matrix from the reference
sample $\bar{\rho}$, and the true variant correlation matrix for
the two-sample summary data $\rho$ may arise due to at least two
reasons. First, the size of the reference sample may be significantly
lower than the sample size of GWASs, thus allowing more room for sampling
errors. Secondly, while all samples are asssumed to be drawn from
the same population, in practice the two correlation estimates may
be based on heterogeneous samples.

To study the problem of mismeasured variant correlations, for any
distinct variants $j,k$, we assume the variant correlation estimate
available to the researcher satisfies $\bar{\rho}_{jk}=\rho_{jk}+\kappa_{jk}$,
where $\rho_{jk}$ is the true variant correlation used to construct
the two-sample summary associations, and $\kappa_{jk}$ is a fixed
effect generated as $\kappa_{jk}\sim\varepsilon\cdot U[0,\bar{\kappa}]$,
$\bar{\kappa}\geq0$, where $\varepsilon$ is generated from the Rademacher
distribution (it equals 1 with 0.5 probability, and -1 with 0.5 probability). 

Figure 4 shows that F-CLR and CLR-01 were best able to control type
I errors under mismeasured variant correlations. S-LIML was slightly
more over-sized, and CLR-20 much more so, especially under large misspecification
($\bar{\kappa}=0.15$). We also note that CLR tests using pruned variants
under correlation thresholds greater than $R^{2}>0.2$ were very unstable.
In a polygenic MR setting, \citet{Wang}'s analysis suggests that
using uncorrelated variants is likely to be a sensible choice when
the quality of variant correlation estimates is in doubt. While CLR-01
is certainly able to control type I errors, in our situation with
balanced errors (variant correlations are equally likely to under
or over-estimated), the S-LIML test and, in particular, the F-CLR
test may have significant power advantages. 
\begin{center}
\includegraphics[width=16.5cm]{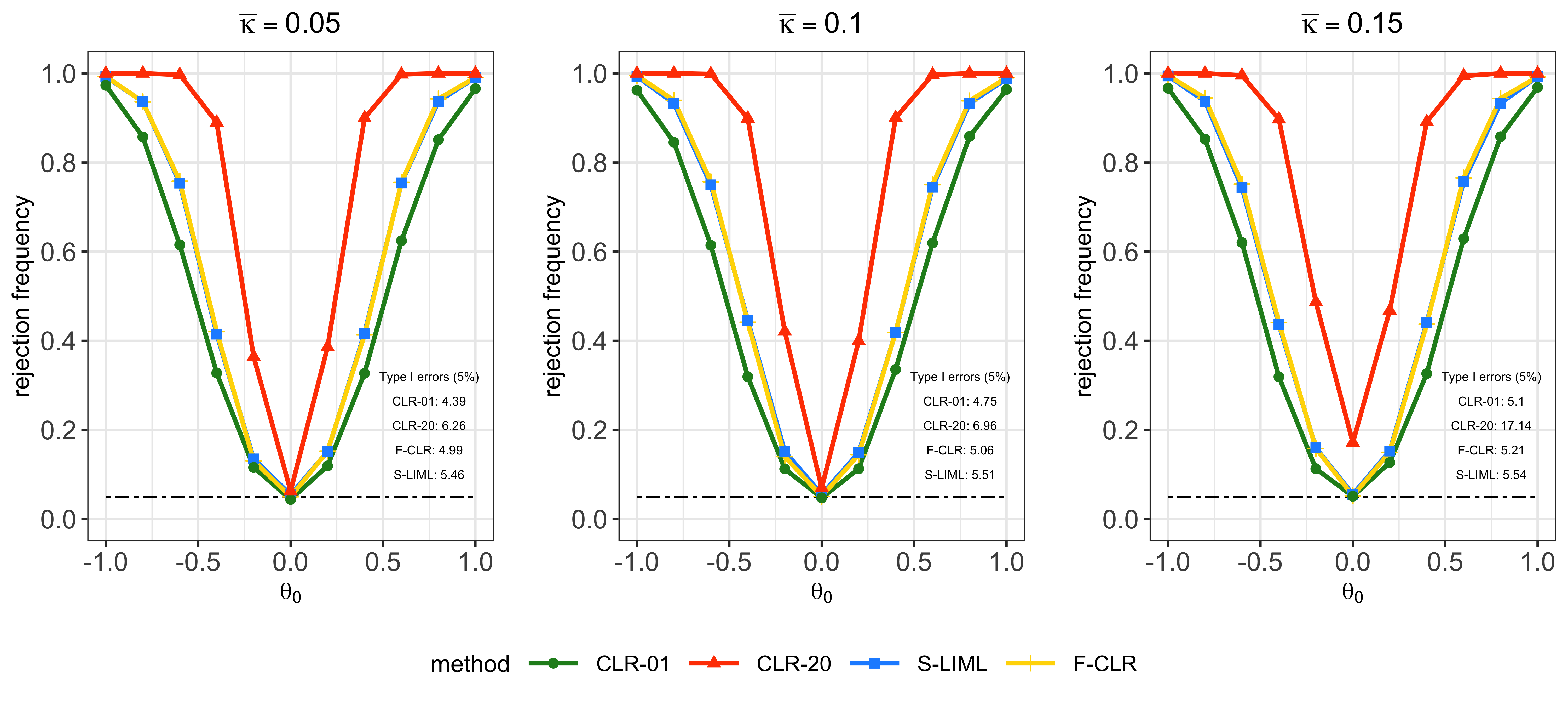}{\footnotesize{}}\\
{\footnotesize{}Figure 4. Power with mismeasured variant correlations
when testing the null hypothesis ${\cal H}_{0}:\theta_{0}=0$. }{\footnotesize\par}
\par\end{center}

Overall, we believe that the S-LIML and F-CLR tests provide a useful
balance between alternative pruning-based approaches; offering robust
inferences in several scenarios of practical concern, without compromising
too much on power. CLR tests with uncorrelated variants appear to
be less powerful than S-LIML and F-CLR tests, and may have poor performance
when variants have fixed and balanced direct effects. On the other
hand, CLR tests which use individual correlated variants as instruments
can be highly sensitive to the pruning threshold chosen, and while
they can be very powerful, their performance may degrade badly under
misspecification. 

\section{Empirical application: CETP inhibition and CHD}

Cholesteryl ester transfer protein (CETP) inhibitors are a class of
drug which increase high-density lipoprotein cholesterol and decrease
low-density lipoprotein cholesterol (LDL-C) concentrations. At least
three CETP inhibitors have failed to provide sufficient evidence of
a protective effect on coronary heart disease (CHD) in clinical trials,
before the successful trial of Anacetrapib showed marginal benefits
alongside statin therapy \citep{Bowman2017}. However, with further
trials currently ongoing, \textsl{cis}-MR analyses can offer important
supporting evidence to complement experimental results. For example,
in recent work, \citet{Schmidt2021}'s \textsl{cis}-MR analysis suggests
that CETP inhibition may be an effective drug target for CHD prevention.

From a statistical perspective, we may have a few concerns regarding
the criteria used by \citet{Schmidt2021} to select instruments. First,
to guard against weak instrument bias, they select variants based
on an in-sample measure of instrument strength (F-statistic > 15),
which could potentially leave the analysis vulnerable to a winner's
curse bias \citep{Mounier2021}. Secondly, to guard against heterogeneity
of genetic associations, they use a measure of instrument validity
to remove outliers (Cochran's Q statistic; see, for example, \citealp{Bowden2019}),
which can result in size-distorted tests \citep{Guggenberger2012}.
Finally, for those correlated variants with strong measured associations
with the outcome, they allow variants up to a pruning threshold of
$R^{2}=0.4$; our simulation results show that inference can be sensitive
to the choice of a pruning threshold.

Here we apply conditional inference techniques to investigate the
genetically-predicted LDL-C lowering effect of CETP inhibition on
the risk of CHD. Genetic associations with LDL-C were taken from a
GWAS of 361,194 individuals of white-British genetic ancestry in the
UK Biobank and were in standard deviation units \citep{Sudlow2015}.
Genetic associations with CHD, measured in log odds ratio (LOR) units,
were taken from a meta-GWAS of 48 studies with a total of 60,801 cases
and 123,504 controls from a majority European population, conducted
by the CARDIoGRAMplusC4D consortium \citep{Consortium2015}. Genetic
variant correlations were obtained from a reference panel of European
individuals \citep[1000 Genomes Project,][]{Auton2015} using the
\texttt{twosampleMR} R package \citep{Hemani2020}. 
\begin{center}
\includegraphics[width=16.5cm]{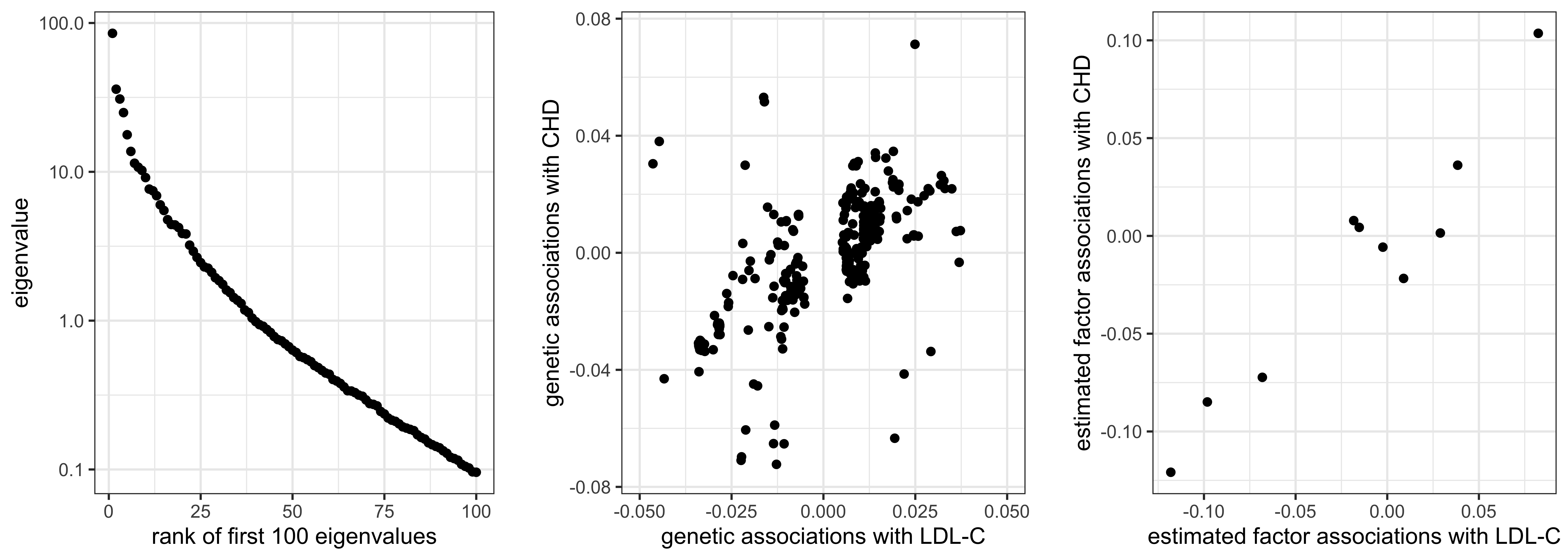}\\
{\footnotesize{}Figure 5. Scree plot (left), 368 genetic associations
with LDL-C and CHD (center), and 10 estimtaed factor associations
with LDL-C and CHD (right), in the }\textsl{\footnotesize{}CETP}{\footnotesize{}
gene region. }{\footnotesize\par}
\par\end{center}

A total of 368 genetic variants were drawn from the \textsl{CETP}
region, with variant positions within $\pm100$ kb from the \textsl{CETP}
gene position indicated on GeneCards \citep{Stelzer2016}. The variant
correlation matrix was highly structured, with 10 principal components
explaining over 96 percent of the total variation of the 368 genetic
variants. Noting the eigenvalue gap between the 10-th and 11-th eigenvalue,
we selected $r=10$ estimated factors as instruments for the F-AR,
F-LM, F-CLR, and S-LIML methods. We also present results for CLR tests
which use individual variants as instruments according to different
pruning thresholds. As in the simulation study, CLR-01, CLR-20, CLR-40,
CLR-60 and CLR-80 denote the CLR test which use variants up to pruning
thresholds $R^{2}=0.01$, $R^{2}=0.2$, $R^{2}=0.4$, $R^{2}=0.6$,
and $R^{2}=0.8$, respectively.

Only 4 of the 10 estimated factors were retained by the S-LIML method
after pre-testing for relevant factors at the $\delta=0.01$ level.
Table 1 shows that the S-LIML method gives a point estimate of $\hat{\theta}_{{\cal S}}=1.075$
for the LOR change in CHD associated with a 1 standard deviation change
in LDL-C, with a corresponding 95 percent asymptotic confidence interval
$\theta_{0}\in[0.245,1.855]$. The results were reasonably robust
to the choice of pre-testing threshold used to select relevant estimated
factors. In particular, pre-tests at the $\delta=0.1$ and $\delta=0.05$
levels resulted in 6 estimated factors being selected, and returned
95 percent confidence intervals of $\theta_{0}\in[0.292,1.801]$ and
$\theta_{0}\in[0.245,1.816]$, respectively. The F-LIML estimator
which uses all 10 estimated factors offers a tighter interval than
the conditional approaches, although its type I error rate can be
more sensitive to weak instruments.
\begin{center}
{\scriptsize{}}%
\begin{tabular}{c||c|c|c|c|c||c|c|c|c|c}
 & {\scriptsize{}F-LIML} & {\scriptsize{}S-LIML} & {\scriptsize{}F-AR} & {\scriptsize{}F-LM} & {\scriptsize{}F-CLR} & {\scriptsize{}CLR-01} & {\scriptsize{}CLR-20} & {\scriptsize{}CLR-40} & {\scriptsize{}CLR-60} & {\scriptsize{}CLR-80}\tabularnewline
\hline 
{\scriptsize{}$\hat{\theta}$} & {\footnotesize{}0.938} & {\footnotesize{}1.075} & {\footnotesize{}-} & {\footnotesize{}-} & {\footnotesize{}-} & {\footnotesize{}-} & {\footnotesize{}-} & {\footnotesize{}-} & {\footnotesize{}-} & {\footnotesize{}-}\tabularnewline
{\scriptsize{}C.I. lower} & {\footnotesize{}0.273} & {\footnotesize{}0.245} & {\footnotesize{}-0.440} & {\footnotesize{}0.282} & {\footnotesize{}0.266} & {\footnotesize{}0.026} & {\footnotesize{}0.565} & {\footnotesize{}0.678} & {\footnotesize{}NA} & {\footnotesize{}0.082}\tabularnewline
{\scriptsize{}C.I. upper} & {\footnotesize{}1.602} & {\footnotesize{}1.855} & {\footnotesize{}2.553} & {\footnotesize{}1.642} & {\footnotesize{}1.661} & {\footnotesize{}1.095} & {\footnotesize{}0.988} & {\footnotesize{}0.942} & {\footnotesize{}NA} & {\footnotesize{}0.152}\tabularnewline
{\scriptsize{}$Q$} & {\footnotesize{}0.999} & {\footnotesize{}0.981} & {\footnotesize{}0.999{*}} & {\footnotesize{}0.999{*}} & {\footnotesize{}0.999{*}} & {\footnotesize{}0.085{*}} & {\footnotesize{}0.205{*}} & {\footnotesize{}0.006{*}} & {\footnotesize{}NA} & {\footnotesize{}0.000{*}}\tabularnewline
\end{tabular}{\scriptsize{}}\\
{\footnotesize{}~}\\
{\footnotesize{}Table 1. 95 percent confidence intervals for pruning
and factor-based approaches. $\hat{\theta}$ gives the point estimate
of the method if applicable. $Q$ gives the p-value associated with
testing the null of no heterogeneity in instrument--LDL-C and instrument--CHD
associations using Cochran's $Q$-statistic; see the discussion below.}{\footnotesize\par}
\par\end{center}

Table 1 also shows that the 95 percent confidence intervals obtained
by inverting the F-CLR and F-LM tests are also tighter than the S-LIML
method, while the F-AR approach is much less precise and unable to
reject the null hypothesis of no causal association (F-AR p-value:
0.508). For CLR tests with pruned variants, although the confidence
intervals are considerably narrower than the factor-based methods,
they appear to be sensitive to the correlation threshold chosen. For
the case of $R^{2}=0.6$, the CLR test rejected all plausible values
of $\theta_{0}$ and so failed to return a confidence interval. 

Our simulation study illustrated how our factor-based approach was
relatively robust to biases from direct variant effects on the outcome.
The heterogeneity plots in Figure 5 can provide a limited insight
into a potential invalid instruments problem. We note that while some
individual variants deviate quite far from the trend line in Figure
5 (center), the heterogeneity appears to be reduced when considering
estimated factor associations (Figure 5, right). 

A more formal method to test for excessive heterogeneity uses Cochran's
$Q$-statistic. Table 1 shows that the F-LIML and S-LIML approaches
provide strong evidence of no heterogeneity when using the estimated
factors as instruments. Since identification-robust methods do not
provide point estimates, for the starred entries in the last row of
Table 1 we evaluated the $Q$-statistic at the mid-point of the relevant
confidence interval. There was no `degrees of freedom' correction
for this substitution which should lead to more conservative p-values
(i.e. we are less likely to reject the null of no heterogeneity).
Despite this, pruning-based approaches show evidence of greater heterogeneity
when considering individual variants as instruments, particularly
at the threshold $R^{2}=0.4$ (CLR-40 heterogeneity p-value: 0.006). 

We conclude that the factor-based conditional testing approaches (F-CLR
and S-LIML) provide robust genetic evidence that the LDL-C lowering
effect of CETP inhibitors is associated with a lower risk of CHD. 

\section{Conclusion}

There is an increasing focus on using\textsl{ cis}-MR analyses to
guide drug development; genetic evidence may be crucial to support
novel targets, precision medicine subgroups, and the design of expensive
clinical trials \citep{Gill2021}. While there are many methods available
for robust inferences in polygenic MR using uncorrelated variants,
powerful \textsl{cis}-MR investigations need to make sensible use
of many weak and correlated genetic associations from the gene region
of interest. Using uncorrelated variants as instruments may lead to
imprecise inferences which can be vulnerable to heterogeneity in genetic
associations. On the other hand, using very correlated variants can
result in unstable inferences which are particularly sensitive to
common problems of misspecification, such as mismeasured genetic correlations. 

With the goal of making reliable and powerful \textsl{cis}-MR inferences,
we have developed two conditional testing procedures based on the
use of genetic factors as instrumental variables. The methods offer
a balance between identification-robust testing procedures which use
individual variants as instruments; offering more robust inferences
without compromising too much on power. 

\subsection*{Funding acknowledgements}

Ashish Patel and Paul Newcombe were funded by the UK Medical Research
Council (programme number MC-UU-00002-9). Paul Newcombe also acknowledges
support from the NIHR Cambridge BRC. Stephen Burgess was supported
by Sir Henry Dale Fellowship jointly funded by the Wellcome Trust
and the Royal Society (grant number 204623-Z-16-Z). Dipender Gill
was funded by the Wellcome 4i Program at Imperial College London (award
number 203928-Z-16-Z), the British Heart Foundation Centre of Research
Excellence (RE-18-4-34215) at Imperial College London, and a National
Institute for Health Research Clinical Lectureship at St George\textquoteright s,
University of London (CL-2020-16-001).

{\small{}\bibliographystyle{chicago}
\bibliography{cis_MR}
}{\small\par}

\newpage
\pagenumbering{arabic}% resets `page` counter to 1
\renewcommand*{\thepage}{[S-\arabic{page}]}
\setcounter{equation}{0}
\renewcommand{\theequation}{S.\arabic{equation}}
\setlength{\parindent}{0pt}

\appendix

\part*{{\LARGE{}Supplementary Material}}

\section{Preparatory lemmata}

\subsection{Notation}

{\small{}We use the following abbreviations: `T' denotes the triangle
inequailty; `CH' denotes Chebyshev's inequality; `CS' denotes the
Cauchy-Schwarz inequality; `M' denotes the Markov inequality; `w.p.a.1'
denotes `with probability approaching 1'; `RHS' denotes `right-hand
side'; `LHS denotes `left-hand side'. For any vector or matrix $A$,
$\Vert A\Vert$ denotes the Euclidean norm. }{\small\par}

\subsection{\citet{Bai2003}'s results on factor loadings}

{\small{}Let $n_{Z}$ denote the sample size of the reference sample
used to compute the genetic variant correlation matrix $\rho$. The
reference sample is permitted to overlap with one of the genetic association
studies. We further assume that $n_{Z}$ is $\Theta(n)$ so that,
in our asymptotic analysis, we consider the sample sizes of the genetic
association studies ($n_{X}$ and $n_{Y}$) and any reference sample
($n_{Z}$) to all be increasing at the same rate with the number of
genetic variants $p$. }{\small\par}

{\small{}Let $z_{ki}$ denote the standardised $k-$th genetic variant
for the $i$-th individual, $i\in[n_{Z}]$. Let $z_{i}=(z_{1i},\ldots,z_{pi})^{\prime}$,
and let $z=(z_{1},\ldots,z_{n_{Z}})^{\prime}$ denote the $n_{Z}\times p$
matrix of standardised genotypes. Under \citet{Bai2002}'s approximate
factor model, we have
\[
z=F\Lambda^{\prime}+e,
\]
where $\Lambda=(\lambda_{1},\ldots,\lambda_{p})^{\prime}$ denotes
a $p\times r$ matrix of factor loadings, $F=(F_{1},\ldots,F_{n_{Z}})^{\prime}$
denotes an $n_{Z}\times r$ matrix of factors, and $e=(e_{1},\ldots,e_{n_{Z}})^{\prime}$
denotes a $n_{Z}\times p$ matrix of idiosyncratic errors. }{\small\par}

{\small{}Let the columns of $\bar{\Lambda}$ denote the first $r$
eigenvectors of the $p\times p$ matrix $z^{\prime}z$ multiplied
by $\sqrt{p}$. Note that we can compute $\bar{\Lambda}$ as the eigenvectors
of the sample variant correlation matrix $\hat{\rho}=z^{\prime}z\big/n_{Z}$
since eigenvectors are invariant to scalar multiplication of a matrix.
Let $\bar{F}=(z\bar{\Lambda})(\bar{\Lambda}^{\prime}\bar{\Lambda})^{-1}$,
and $\hat{F}=\bar{F}(\bar{F}^{\prime}\bar{F}\big/n_{Z})^{-\frac{1}{2}}$.
Then, $\hat{F}^{\prime}\hat{F}\big/n_{Z}=I_{r}$, and $\hat{\Lambda}=\hat{F}^{\prime}z\big/n_{Z}=(\bar{F}^{\prime}\bar{F}\big/n_{Z})^{-\frac{1}{2}}(\bar{\Lambda}^{\prime}\bar{\Lambda})^{-1}(\bar{\Lambda}^{\prime}z^{\prime}z\big/n_{Z})$. }{\small\par}

{\small{}Let $H=(\Lambda^{\prime}\Lambda\big/p)(F^{\prime}\hat{F}\big/n_{Z})D_{np}^{-1}$
where $D_{np}$ is an $r\times r$ diagonal matrix with its diagonal
entries equal to the first $r$ largest eigenvalues of $(zz^{\prime}\big/(n_{Z}p))$
in decreasing order. }{\small\par}

\textbf{\small{}Lemma 1. }{\small{}As $n,p\to\infty$,}\textbf{\small{}
(i)}{\small{} $\Vert H\Vert=O_{P}(1)$; }\textbf{\small{}(ii)}{\small{}
$\Vert H^{-1}\Vert=O_{P}(1)$. }{\small\par}

{\small{}Proof. By Lemma A.3(i) of \citet[p.161]{Bai2003}, $D_{np}\overset{P}{\to}D$,
a diagonal matrix consisting of the eigenvalues of $\Sigma_{\Lambda}\Sigma_{F}$.
Hence, w.p.a.1, $D_{np}^{-1}$ exists since $\Sigma_{\Lambda}$ and
$\Sigma_{F}$ are positive definite, and $\Vert D_{np}^{-1}\Vert=O_{P}(1)$.
By construction, $\hat{F}^{\prime}\hat{F}\big/n_{Z}=I_{r}$, so that
$\Vert\hat{F}^{\prime}\hat{F}\big/n_{Z}\Vert=O_{P}(1)$. By CS, Assumption
2, and M, $\Vert F^{\prime}F\big/n_{Z}\Vert\leq\sum_{i=1}^{n_{Z}}\Vert F_{i}\Vert^{2}\big/n_{Z}=O_{P}(1)$.
Similarly, for all $k\in[p]$, by Assumption 2, $\Vert\lambda_{k}\Vert\leq C_{\lambda}<\infty$.
Therefore, $\Vert\Lambda^{\prime}\Lambda\big/p\Vert\leq\sum_{k=1}^{p}\Vert\lambda_{k}\Vert^{2}\big/p=O(1)$.
Also, by CS, $\Vert H\Vert\leq\Vert\Lambda^{\prime}\Lambda\big/p\Vert\cdot\Vert F^{\prime}F\big/n_{Z}\Vert^{\frac{1}{2}}\cdot\Vert\hat{F}^{\prime}\hat{F}\big/n_{Z}\Vert^{\frac{1}{2}}\cdot\Vert D_{np}^{-1}\Vert$,
so that $\Vert H\Vert=O_{P}(1)$. $H$ is invertible; see \citet[p.145]{Bai2003}.
Therefore, by Part (i), $\Vert H^{-1}\Vert=O_{P}(1)$. \hfill{}$\square$}{\small\par}

\textbf{\small{}Lemma 2.}{\small{} As $n,p\to\infty$, $\Vert\hat{\lambda}_{k}-H^{-1}\lambda_{k}\Vert=O_{P}(n^{-\frac{1}{2}})+O_{P}(\min(n,p)^{-1})$,
$k\in[p]$. }{\small\par}

{\small{}Proof. By \citet[p.165]{Bai2003}, for any $k\in[p]$, we
have 
\begin{equation}
\hat{\lambda}_{k}-H^{-1}\lambda_{k}=H^{\prime}\frac{1}{n_{Z}}\sum_{i=1}^{n_{Z}}F_{i}e_{ki}+\frac{1}{n_{Z}}\sum_{i=1}^{n_{Z}}\hat{F}_{i}(F_{i}-H^{-1\prime}\hat{F}_{i})\lambda_{k}+\frac{1}{n_{Z}}\sum_{i=1}^{n_{Z}}(\hat{F}_{i}-H^{\prime}F_{i})e_{ki},
\end{equation}
where $e_{ki}$ is the $(i,k)$-th element of the matrix of idiosyncratic
errors $e$. Hence, by CS, 
\[
\Vert\hat{\lambda}_{k}-H^{-1}\lambda_{k}\Vert=\frac{1}{\sqrt{n_{Z}}}\Vert H\Vert\cdot\Big\Vert\frac{1}{\sqrt{n_{Z}}}\sum_{i=1}^{n_{Z}}F_{i}e_{ki}\Big\Vert+\Big\Vert\frac{1}{n_{Z}}\sum_{i=1}^{n_{Z}}\hat{F}_{i}(F_{i}-H^{\prime-1}\hat{F}_{i})\Big\Vert\cdot\Vert\lambda_{k}\Vert+\Big\Vert\frac{1}{n_{Z}}\sum_{i=1}^{n_{Z}}(\hat{F}_{i}-H^{\prime}F_{i})e_{ki}\Big\Vert,
\]
where the first term on the RHS is $O_{P}(n^{-\frac{1}{2}})$ by Lemma
1(i), Assumption D of \citet[p.141]{Bai2003}, and M. The second term
on the RHS is $O_{P}(\min(n,p)^{-1})$ by Lemma B.3 of \citet[p.165]{Bai2003},
and Assumption 2. The third term on the RHS is $O_{P}(\min(n,p)^{-1})$
by Lemma B.1 of \citet[p.163]{Bai2003}. \hfill{}$\square$}{\small\par}

\subsection{Assumption S0: weak dependence of model errors}

{\small{}Under our model assumptions, $\beta_{Y}=\beta_{X}\theta_{0}$.
Using Assumption 1, we can write $\hat{\beta}_{Y}-\hat{\beta}_{X}\theta_{0}=\varepsilon_{Y}-\varepsilon_{X}\theta_{0}$,
where $\varepsilon_{X}\sim N(0,\Sigma_{X})$ and $\varepsilon_{Y}\sim N(0,\Sigma_{Y})$.
Let $\varepsilon=\varepsilon_{Y}-\theta_{0}\varepsilon_{X}$, and
define $\varepsilon^{\star}=\sqrt{n}\varepsilon$. Similarly, define
$\varepsilon_{X}^{\star}=\sqrt{n}\varepsilon_{X}$. Note that since
$\varepsilon\sim N(0,\Sigma_{Y}+\theta_{0}^{2}\Sigma_{X})$, and $\Sigma_{Y}+\theta_{0}^{2}\Sigma_{X}=\Theta(n^{-1})$,
the re-scaled errors satisfy $\varepsilon^{\star}=O(1)$ and $\varepsilon_{X}^{\star}=O(1)$.
We maintain the following assumptions. }{\small\par}

\textbf{\small{}Assumption S0.}{\small{} There exists a universal
positive constant $C$, such that, as $n_{Z},p\to\infty$,}\textbf{\small{}
(i)}{\small{} for each $i\in[n_{Z}]$, $\sum_{k=1}^{p}\vert\mathbb{E}[e_{ki}\varepsilon_{k}^{\star}]\vert\leq C$
and $\sum_{k=1}^{p}\vert\mathbb{E}[e_{ki}\varepsilon_{X_{k}}^{\star}]\vert\leq C$;
}\textbf{\small{}(ii)}{\small{} $\mathbb{E}\big[\Vert\frac{1}{\sqrt{n_{Z}p}}\sum_{i=1}^{n_{Z}}\sum_{k=1}^{p}F_{i}(e_{ki}\varepsilon_{k}^{\star}-\mathbb{E}[e_{ki}\varepsilon_{k}^{\star}])\Vert\big]\leq C$
and $\mathbb{E}\big[\Vert\frac{1}{\sqrt{n_{Z}p}}\sum_{i=1}^{n_{Z}}\sum_{k=1}^{p}F_{i}(e_{ki}\varepsilon_{X_{k}}^{\star}-\mathbb{E}[e_{ki}\varepsilon_{X_{k}}^{\star}])\Vert\big]\leq C$;
}\textbf{\small{}(iii)}{\small{} for each $i\in[n_{Z}]$, $\mathbb{E}\big[\vert\frac{1}{\sqrt{p}}\sum_{k=1}^{p}(e_{ki}\varepsilon_{k}^{\star}-\mathbb{E}[e_{ki}\varepsilon_{k}^{\star}])\vert^{2}\big]\leq C$
and $\mathbb{E}\big[\vert\frac{1}{\sqrt{p}}\sum_{k=1}^{p}(e_{ki}\varepsilon_{X_{k}}^{\star}-\mathbb{E}[e_{ki}\varepsilon_{X_{k}}^{\star}])\vert^{2}\big]\leq C$;
}\textbf{\small{}(iv) }{\small{}for each $i\in[n_{Z}]$, $\mathbb{E}\big[\vert\frac{1}{\sqrt{p}}\sum_{k=1}^{p}e_{ki}\vert^{2}\big]\leq C$
and $\mathbb{E}\big[\Vert\frac{1}{\sqrt{n_{Z}p}}\sum_{i=1}^{n_{Z}}\sum_{k=1}^{p}F_{i}e_{ki}\Vert\big]\leq C$. }{\small\par}

{\small{}The conditions in Assumption S0 are similar to \citet{Bai2010}.
Assumption S0(i) restricts the extent to which idiosyncratic errors
of the approximate factor model can be correlated with sampling errors
of genetic associations, $\varepsilon_{X}=\hat{\beta}_{X}-\beta_{X}$
and $\varepsilon_{Y}=\hat{\beta}_{Y}-\beta_{Y}$. This assumption
is trivially satisfied when the reference sample used to be obtain
genetic correlations is independent of the both genetic association
studies, as in our real data example in Section 5. For Parts (ii),
(iii) and (iv), we note that these are conditions on zero-mean sums,
and are analogous to Assumptions C5, D, F1, and F2 of \citet[p.141 and p.144]{Bai2003}. }{\small\par}

\subsection{Covariance terms }

{\small{}Let $\Omega_{X}=H^{-1}\Lambda^{\prime}\Sigma_{X}\Lambda H^{-1\prime}$,
$\Omega_{Y}=H^{-1}\Lambda^{\prime}\Sigma_{Y}\Lambda H^{-1\prime}$,
$\hat{\Omega}_{X}=\hat{\Lambda}^{\prime}\Sigma_{X}\hat{\Lambda}$,
and $\hat{\Omega}_{Y}=\hat{\Lambda}^{\prime}\Sigma_{Y}\hat{\Lambda}$.
Note that $\hat{\Omega}(\theta)=\hat{\Omega}_{Y}+\theta^{2}\hat{\Omega}_{X}$,
and let $\Omega=\Omega_{Y}+\theta_{0}^{2}\Omega_{X}$. }{\small\par}

\textbf{\small{}Lemma 3.}{\small{} For any consistent estimator $\dot{\theta}$
of $\theta_{0}$, as $n,p\to\infty$, }\textbf{\small{}(i)}{\small{}
$\Omega_{X}=O_{P}(1)$ and $\Omega=O_{P}(1)$; }\textbf{\small{}(ii)}{\small{}
$\Omega_{X}^{-1}=O_{P}(1)$ and $\Omega^{-1}=O_{P}(1)$; }\textbf{\small{}(iii)}{\small{}
$\hat{\Omega}_{X}-\Omega_{X}=o_{P}(1)$ and $\hat{\Omega}(\dot{\theta})-\Omega=o_{P}(1)$;
}\textbf{\small{}(iv)}{\small{} $\hat{\Omega}_{X}=O_{P}(1)$ and $\hat{\Omega}(\dot{\theta})=O_{P}(1)$;
}\textbf{\small{}(v)}{\small{} $\hat{\Omega}_{X}^{-1}=O_{P}(1)$ and
$\hat{\Omega}(\dot{\theta})^{-1}=O_{P}(1)$; }\textbf{\small{}(vi)}{\small{}
$\hat{\Omega}_{X}^{-1}-\Omega_{X}^{-1}=o_{P}(1)$ and $\hat{\Omega}(\dot{\theta})^{-1}-\Omega^{-1}=o_{P}(1)$. }{\small\par}

{\small{}Proof. Since $\Lambda^{\prime}\Lambda=O(p)$ by Assumption
2, we have that $\Lambda^{\prime}\Sigma_{X}\Lambda=\Theta(n^{-1}\Lambda^{\prime}\Lambda)=O(n^{-1}p)$
by Assumption 1. Similarly, $\Lambda^{\prime}\Sigma_{X}\Lambda=O(n^{-1}p)$.
Part (i) then follows by Lemma 1(ii). Moreover, $\Lambda^{\prime}\Sigma_{X}\Lambda$
and $\Lambda^{\prime}(\Sigma_{Y}+\theta_{0}^{2}\Sigma_{X})\Lambda$
are invertible, so that Part (ii) follows by Lemma 1(i). For Part
(iii), note that $\hat{\Omega}_{X}-\Omega_{X}=(\hat{\Lambda}-\Lambda H^{-1\prime})^{\prime}\Sigma_{X}(\hat{\Lambda}-\Lambda H^{-1\prime})+2(\hat{\Lambda}-\Lambda H^{-1\prime})^{\prime}\Sigma_{X}\Lambda H^{-1\prime}$.
Then,
\begin{eqnarray*}
\Vert\hat{\Omega}_{X}-\Omega_{X}\Vert & \leq & \Theta(n^{-1})\sum_{k=1}^{p}\Vert\hat{\lambda}_{k}-H^{-1}\lambda_{k}\Vert^{2}+\Theta(n^{-1})\Vert H^{-1}\Vert\sum_{k=1}^{p}\Vert\hat{\lambda}_{k}-H^{-1}\lambda_{k}\Vert\cdot\Vert\lambda_{k}\Vert\\
 & = & O_{P}(n^{-\frac{3}{2}}p)+O_{P}(n^{-1}\min(n,p)^{-1}p)\\
 & = & o_{P}(1),
\end{eqnarray*}
by CS, T, Assumptions 1 and 2, and Lemmas 1(ii) and 2. By identical
arguments, $\Vert\hat{\Omega}_{Y}-\Omega_{Y}\Vert=o_{P}(1)$. Thus,
by CS, T, consistency of $\dot{\theta}$, and Part (i), $\Vert\hat{\Omega}(\dot{\theta})-\Omega\Vert\leq\Vert\hat{\Omega}_{Y}-\Omega_{Y}\Vert+\dot{\theta}^{2}\Vert\hat{\Omega}_{X}-\Omega_{X}\Vert+\vert\dot{\theta}-\theta_{0}\vert\vert\dot{\theta}+\theta_{0}\vert\Vert\Omega_{X}\Vert=o_{P}(1)$.
Part (iv) then follows by Parts (i), (iii), and T. By Part (iii),
$np^{-1}(\hat{\Omega}_{X}-\Omega_{X})=o_{P}(1)$, so that, by Part
(i), $np^{-1}\hat{\Omega}_{X}$ is invertible w.p.a.1, and $\hat{\Omega}_{X}^{-1}=O_{P}(1)$.
Similarly, $\hat{\Omega}(\dot{\theta})^{-1}=O_{P}(1)$. Finally, $\hat{\Omega}_{X}^{-1}-\Omega_{X}^{-1}=\hat{\Omega}_{X}^{-1}(\Omega_{X}-\hat{\Omega}_{X})\Omega_{X}^{-1}=O_{P}(1)o_{P}(1)=o_{P}(1)$,
by Parts (ii), (iii), and (v). Identical arguments show that $\hat{\Omega}(\dot{\theta})^{-1}-\Omega^{-1}=o_{P}(1)$.
\hfill{}$\square$}{\small\par}

\subsection{Derivative terms}

{\small{}Let $\hat{G}=-\hat{\Lambda}^{\prime}\hat{\beta}_{X}$, $G=-H^{-1}\Lambda^{\prime}\beta_{X}$,
and $\varepsilon_{X}\sim N(0,\Sigma_{X})$. }{\small\par}

\textbf{\small{}Lemma 4. }{\small{}As $n,p\to\infty$, }\textbf{\small{}(i)}{\small{}
$\hat{G}-G=O_{P}(1)$; }\textbf{\small{}(ii)}{\small{} $\hat{G}=O_{P}(p^{\frac{1}{2}})$.}{\small\par}

{\small{}Proof. We can write $\hat{G}-G=-(\hat{\Lambda}-\Lambda H^{-1\prime})^{\prime}(\beta_{X}+\varepsilon_{X})-H^{-1}\Lambda^{\prime}\varepsilon_{X}$.
By CS and T, 
\[
\Vert(\hat{\Lambda}-\Lambda H^{-1\prime})^{\prime}(\beta_{X}+\varepsilon_{X})\Vert\leq\Big(\sum_{k=1}^{p}\Vert\hat{\lambda}_{k}-H^{-1}\lambda_{k}\Vert^{2}\Big)^{\frac{1}{2}}\big(\Vert\beta_{X}\Vert+\Vert\varepsilon_{X}\Vert\big).
\]
Note that $\Vert\beta_{X}\Vert=O(1)$ by Assumption 3, and $\mathbb{E}[\Vert\varepsilon_{X}\Vert]\leq\big(\sum_{k=1}^{p}\mathbb{E}[\vert\varepsilon_{X_{k}}\vert^{2}]\big)^{\frac{1}{2}}=O_{P}(n^{-\frac{1}{2}}p^{\frac{1}{2}})$
by CS and Assumption 1. Hence, by Lemma 2 and M, the RHS of the above
equation is $O_{P}(n^{-\frac{1}{2}}p^{\frac{1}{2}})+O_{P}(\min(n,p)^{-1}p^{\frac{1}{2}})+O_{P}(n^{-1}p^{-1})+O_{P}(n^{-\frac{1}{2}}\min(n,p)^{-1}p)$.
Also, since $Var(H^{-1}\Lambda^{\prime}\varepsilon_{X})=\Omega_{X}$,
we have that $H^{-1}\Lambda^{\prime}\varepsilon_{X}=O_{P}(n^{-\frac{1}{2}}p^{\frac{1}{2}})$
by CH and Lemma 3(i). For Part (ii), $\Lambda^{\prime}\beta_{X}=\Theta(\sqrt{p})$
under a strong instruments assumption. The result then follows by
Part (i) and T. \hfill{}$\square$ }{\small\par}

\subsection{Residual term}

{\small{}We can write $\hat{\beta}_{Y}-\hat{\beta}_{X}\theta_{0}=\varepsilon_{Y}-\varepsilon_{X}\theta_{0}$,
where $\varepsilon_{X}\sim N(0,\Sigma_{X})$ and $\varepsilon_{Y}\sim N(0,\Sigma_{Y})$.
Recall from Section 1.3, $\varepsilon=\varepsilon_{Y}-\theta_{0}\varepsilon_{X}$,
and $\varepsilon^{\star}=\varepsilon\sqrt{n}$. Let $\hat{g}(\theta)=\hat{\Lambda}^{\prime}(\hat{\beta}_{Y}-\hat{\beta}_{X}\theta)$
and $g(\theta)=H^{-1}\Lambda^{\prime}(\hat{\beta}_{Y}-\hat{\beta}_{X}\theta)$. }{\small\par}

\textbf{\small{}Lemma 5. }{\small{}As $n,p\to\infty$, }\textbf{\small{}(i)}{\small{}
$\hat{g}(\theta_{0})-g(\theta_{0})=o_{P}(1)$; }\textbf{\small{}(ii)}{\small{}
$\hat{g}(\theta_{0})=O_{P}(1)$; }\textbf{\small{}(iii)}{\small{}
for any consistent estimator $\dot{\theta}$ of $\theta_{0}$, $\hat{g}(\dot{\theta})=o_{P}(p^{\frac{1}{2}})$. }{\small\par}

{\small{}Proof. We can write
\begin{eqnarray*}
\hat{g}(\theta_{0})-g(\theta_{0}) & = & \sum_{k=1}^{p}(\hat{\lambda}_{k}-H^{-1}\lambda_{k})\varepsilon_{k}\\
 & = & H^{\prime}\frac{1}{n_{Z}}\sum_{i=1}^{n_{Z}}\sum_{k=1}^{p}F_{i}e_{ki}\varepsilon_{k}+\frac{1}{n_{Z}}\sum_{i=1}^{n_{Z}}\sum_{k=1}^{p}\hat{F}_{i}(F_{i}-H^{-1\prime}\hat{F}_{i})^{\prime}\lambda_{k}\varepsilon_{k}\\
 &  & +\frac{1}{n_{Z}}\sum_{i=1}^{n_{Z}}\sum_{k=1}^{p}(\hat{F}_{i}-H^{\prime}F_{i})e_{ki}\varepsilon_{k}\\
 & := & H^{\prime}{\cal R}_{1}+{\cal R}_{2}+{\cal R}_{3}
\end{eqnarray*}
}{\small\par}

{\small{}First, note that 
\[
\sqrt{n}{\cal R}_{1}=\sqrt{\frac{p}{n_{Z}}}\sqrt{\frac{1}{n_{Z}p}}\sum_{k=1}^{p}\sum_{i=1}^{n_{Z}}F_{i}(e_{ki}\varepsilon_{k}^{\star}-\mathbb{E}[e_{ki}\varepsilon_{k}^{\star}])+\frac{1}{n_{Z}}\sum_{i=1}^{n_{Z}}F_{i}\sum_{k=1}^{p}\mathbb{E}[e_{ki}\varepsilon_{k}^{\star}].
\]
By CS, 
\begin{eqnarray*}
\sqrt{n}\Vert{\cal R}_{1}\Vert & \leq & \sqrt{\frac{p}{n_{Z}}}\Big\Vert\frac{1}{\sqrt{n_{Z}p}}\sum_{k=1}^{p}\sum_{i=1}^{n_{Z}}F_{i}(e_{ki}\varepsilon_{k}^{\star}-\mathbb{E}[e_{ki}\varepsilon_{k}^{\star}])\Big\Vert+\Big(\frac{1}{n_{Z}}\sum_{i=1}^{n_{Z}}\Vert F_{i}\Vert^{2}\Big)^{\frac{1}{2}}\Big(\frac{1}{n_{Z}}\sum_{i=1}^{n_{Z}}\Big(\sum_{k=1}^{p}\vert\mathbb{E}[e_{ki}\varepsilon_{k}^{\star}]\vert\Big)^{2}\Big)^{\frac{1}{2}}\\
 & = & O_{P}(1),
\end{eqnarray*}
by Assumption 2, Assumptions S0(i) and (ii), CS, T, and M. Thus, by
Lemma 1(i) and CS, $H^{\prime}{\cal R}_{1}=o_{P}(1)$. }{\small\par}

{\small{}For ${\cal R}_{2}$, by CS, Lemma 1(ii), and Lemma B.3 of
\citet[p.165]{Bai2003}, 
\begin{eqnarray*}
\Big\Vert\frac{1}{n_{Z}}\hat{F}^{\prime}(F-\hat{F}H^{-1})\Big\Vert & \leq & \Big\Vert\frac{1}{n_{Z}}\hat{F}^{\prime}(FH-\hat{F})\Big\Vert\cdot\Vert H^{-1}\Vert\\
 & = & O_{P}(\min(n,p)^{-1}).
\end{eqnarray*}
Also, $\Vert\Lambda^{\prime}\varepsilon\Vert=O_{P}(p^{\frac{1}{2}}n^{-\frac{1}{2}})$
by CH and Assumptions 1 and 2. Hence, by CS, $\Vert{\cal R}_{2}\Vert=O_{P}(p^{\frac{1}{2}}n^{-\frac{1}{2}})$
$\times O_{P}(\min(n,p)^{-1})=o_{P}(1)$. }{\small\par}

{\small{}Finally, for ${\cal R}_{3}$, note that 
\[
\sqrt{n}{\cal R}_{3}=\frac{1}{n_{Z}}\sum_{i=1}^{n_{Z}}(\hat{F}_{i}-H^{\prime}F_{i})\sum_{k=1}^{p}(e_{ki}\varepsilon_{k}^{\star}-\mathbb{E}[e_{ki}\varepsilon_{k}^{\star}])+\frac{1}{n_{Z}}\sum_{i=1}^{n_{Z}}(\hat{F}_{i}-H^{\prime}F_{i})\sum_{k=1}^{p}\mathbb{E}[e_{ki}\varepsilon_{k}^{\star}].
\]
Then, by CS, 
\begin{eqnarray*}
\Vert{\cal R}_{3}\Vert & \leq & \sqrt{\frac{p}{n}}\Big(\frac{1}{n_{Z}}\sum_{i=1}^{n_{Z}}\Vert\hat{F}_{i}-H^{\prime}F_{i}\Vert^{2}\Big)^{\frac{1}{2}}\Big(\frac{1}{n_{Z}}\sum_{i=1}^{n_{Z}}\Big\vert\frac{1}{\sqrt{p}}\sum_{k=1}^{p}(e_{ki}\varepsilon_{k}^{\star}-\mathbb{E}[e_{ki}\varepsilon_{k}^{\star}])\Big\vert^{2}\Big)^{\frac{1}{2}}\\
 &  & +\frac{1}{\sqrt{n}}\Big(\frac{1}{n_{Z}}\sum_{i=1}^{n_{Z}}\Vert\hat{F}_{i}-H^{\prime}F_{i}\Vert^{2}\Big)^{\frac{1}{2}}\Big(\frac{1}{n_{Z}}\sum_{i=1}^{n_{Z}}\big(\sum_{k=1}^{p}\vert\mathbb{E}[e_{ki}\varepsilon_{k}^{\star}]\vert\big)^{2}\Big)^{\frac{1}{2}}\\
 & = & O_{P}(\min(n,p)^{-\frac{1}{2}})\\
 & = & o_{P}(1),
\end{eqnarray*}
by M, T, Lemma A.1 of \citet[p.159]{Bai2003}, and Assumptions S0(ii)
and (iii). By the above results on the remainder terms ${\cal R}_{1}$,
${\cal R}_{2}$, and ${\cal R}_{3}$, Part (i) follows by T. }{\small\par}

{\small{}For Part (ii), note that $g(\theta_{0})\sim N(0,\Omega)$,
so that by CH and Lemma 3(i), $g(\theta_{0})=O_{P}(1)$. Hence, by
Part (i) and T, $\hat{g}(\theta_{0})=O_{P}(1)$. Similarly, $\hat{g}(\dot{\theta})-\hat{g}(\theta_{0})=\hat{G}(\dot{\theta}-\theta_{0})=o_{P}(p^{\frac{1}{2}})$
by Lemma 4(ii), CS, and consistency of $\dot{\theta}$ for $\theta_{0}$.
Hence, Part (iii) follows by Part (ii) and T. \hfill{}$\square$ }{\small\par}

\section{Proof of Theorem 1: F-LIML with strong instruments}

{\small{}The F-LIML estimator is given by $\hat{\theta}_{F}=\arg\max_{\theta}\hat{Q}(\theta)$,
where $\hat{Q}(\theta)=-\hat{g}(\theta)^{\prime}\hat{\Omega}(\theta)^{-1}\hat{g}(\theta)\big/2$.
Under the conditions of Theorem 1, the estimator $\hat{\theta}_{F}$
can be shown to be consistent for $\theta_{0}$ by applying standard
arguments used to establish consistency of extremum estimators; see,
for example, \citet[Theorem 2.1, p.2121]{Newey1994a} and \citet[Proof of Theorem 3.1]{Zhao2018}.
By a first-order Taylor expansion, there exists a $\dot{\theta}$
in the line segment joining $\hat{\theta}_{F}$ and $\theta_{0}$
such that $\nabla_{\theta\theta}\hat{Q}(\dot{\theta})(\hat{\theta}_{F}-\theta_{0})=-\nabla_{\theta}\hat{Q}(\theta_{0})$.
Dividing both sides by $-G^{\prime}\Omega^{-1}G$, and by Lemmas 7(i)
and (ii) below, it follows that $(G^{\prime}\Omega^{-1}G)^{\frac{1}{2}}(\hat{\theta}_{F}-\theta_{0})\overset{D}{\to}N(0,1)$.
Let $\hat{\Omega}=\hat{\Omega}(\hat{\theta}_{F})$. Now, since $\hat{G}^{\prime}\hat{\Omega}^{-1}\hat{G}-G^{\prime}\Omega^{-1}G=\hat{G}^{\prime}(\hat{\Omega}^{-1}-\Omega^{-1})\hat{G}+(\hat{G}-G)^{\prime}\Omega^{-1}\hat{G}+G^{\prime}\Omega^{-1}(\hat{G}-G)=o_{P}(n)$
by Lemmas 3 and 4. Therefore, $(G^{\prime}\Omega^{-1}G)^{-1}(\hat{G}^{\prime}\hat{\Omega}^{-1}\hat{G})=o_{P}(1)$,
and the result of the theorem follows by Slutsky's lemma. }{\small\par}

\textbf{\small{}Lemma 6.}{\small{} Under $p=\Theta(n)$, as $n,p\to\infty$,}\textbf{\small{}
(i)}{\small{} $\nabla_{\theta}\hat{Q}(\theta_{0})=-G^{\prime}\Omega{}^{-1}\hat{g}(\theta_{0})+o_{P}(n^{\frac{1}{2}})+O_{P}(1)$;
}\textbf{\small{}(ii)}{\small{} $\nabla_{\theta\theta}\hat{Q}(\dot{\theta})=-G^{\prime}\Omega^{-1}G+O_{P}(n^{\frac{1}{2}})+o_{P}(n)$. }{\small\par}

{\small{}Proof. For Part (i), we have 
\begin{eqnarray*}
\nabla_{\theta}\hat{Q}(\theta_{0}) & = & -\hat{G}^{\prime}\hat{\Omega}^{-1}\hat{g}(\theta_{0})+\theta_{0}\hat{g}(\theta_{0})^{\prime}\hat{\Omega}_{X}\hat{\Omega}^{-2}\hat{g}(\theta_{0})\\
 & = & -(\hat{G}-G)^{\prime}\hat{\Omega}^{-1}\hat{g}(\theta_{0})-G^{\prime}(\hat{\Omega}^{-1}-\Omega^{-1})\hat{g}(\theta_{0})-G^{\prime}\Omega{}^{-1}\hat{g}(\theta_{0})\\
 &  & +\theta_{0}\hat{g}(\theta_{0})^{\prime}(\hat{\Omega}_{X}-\Omega_{X})\hat{\Omega}^{-2}\hat{g}(\theta_{0})+\theta_{0}\hat{g}(\theta_{0})^{\prime}\Omega_{X}(\hat{\Omega}^{-2}-\Omega^{-2})\hat{g}(\theta_{0})+\theta_{0}\hat{g}(\theta_{0})^{\prime}\Omega_{X}\Omega^{-2}\hat{g}(\theta_{0})\\
 & = & -G^{\prime}\Omega{}^{-1}\hat{g}(\theta_{0})+o_{P}(n^{\frac{1}{2}})+O_{P}(1),
\end{eqnarray*}
by Lemmas 3, 4, and 5(ii), and noting that $\hat{\Omega}^{-2}-\Omega^{-2}=(\hat{\Omega}^{-1}-\Omega^{-1})(\hat{\Omega}^{-1}+\Omega^{-1})=o_{P}(1)$
by Lemmas 3(ii), (v), and (vi).}{\small\par}

{\small{}Similarly, for Part (ii), 
\begin{eqnarray*}
\nabla_{\theta\theta}\hat{Q}(\dot{\theta}) & = & -\hat{G}^{\prime}\hat{\Omega}(\dot{\theta})^{-1}\hat{G}+\hat{g}(\dot{\theta})^{\prime}\hat{\Omega}_{X}\hat{\Omega}(\dot{\theta})^{-2}\hat{g}(\dot{\theta})+4\dot{\theta}\hat{G}^{\prime}\hat{\Omega}_{X}\hat{\Omega}(\dot{\theta})^{-2}\hat{g}(\dot{\theta})-4\dot{\theta}^{2}\hat{g}(\dot{\theta})^{\prime}\hat{\Omega}_{X}^{2}\hat{\Omega}(\dot{\theta})^{-3}\hat{g}(\dot{\theta})\\
 & = & -G^{\prime}\Omega^{-1}G-(\hat{G}-G)^{\prime}\hat{\Omega}(\dot{\theta})^{-1}\hat{G}-G^{\prime}(\hat{\Omega}(\dot{\theta})^{-1}-\Omega^{-1})\hat{G}-G^{\prime}\Omega^{-1}(\hat{G}-G)+o_{P}(p)\\
 & = & -G^{\prime}\Omega^{-1}G+O_{P}(n^{\frac{1}{2}})+o_{P}(n),
\end{eqnarray*}
using Lemmas 3, 4, and 5(iii). \hfill{}$\square$ }{\small\par}

\textbf{\small{}Lemma 7. }{\small{}As $n,p\to\infty$,}\textbf{\small{}
(i)}{\small{} $G^{\prime}\Omega^{-1}G=O(n)$; }\textbf{\small{}(ii)}{\small{}
$(G^{\prime}\Omega^{-1}G)^{-\frac{1}{2}}\nabla_{\theta}\hat{Q}(\theta_{0})\overset{D}{\to}N(0,I_{r})$;
}\textbf{\small{}(iii)}{\small{} $(G^{\prime}\Omega^{-1}G)^{-1}\nabla_{\theta\theta}\hat{Q}(\dot{\theta})\overset{P}{\to}-I_{r}$. }{\small\par}

{\small{}Proof. First, $G=-H^{-1}\Lambda^{\prime}\beta_{X}=\Theta(\sqrt{p})$
by a strong instruments assumption and Lemma 1, so that Part (i) follows
by Lemma 3(ii). Therefore, Part (ii) follows by Lemma 6(i) and Slutsky's
lemma since $(G^{\prime}\Omega^{-1}G)^{-\frac{1}{2}}\nabla_{\theta}\bar{Q}(\theta_{0})=-(G^{\prime}\Omega^{-1}G)^{-\frac{1}{2}}G^{\prime}\Omega{}^{-1}\hat{g}(\theta_{0})+o_{P}(1)\overset{D}{\to}N(0,I_{r})$,
and $\hat{g}(\theta_{0})=g(\theta_{0})+o_{P}(1)\overset{D}{\to}N(0,\Omega)$
by Lemma 5(i). Similarly, for Part (iii), $(G^{\prime}\Omega^{-1}G)^{-1}\nabla_{\theta\theta}\hat{Q}(\dot{\theta})=-I_{r}+o_{P}(1)$
by Part (i) and Lemma 6(ii). \hfill{}$\square$}{\small\par}

\section{Proof of Theorem 2: Identification-robust tests}

{\small{}Under our weak instrument asymptotics, $\Lambda^{\prime}\beta_{X}=\Theta(1)$,
so that $G=-H^{-1}\Lambda^{\prime}\beta_{X}=\Theta(1)$ by Lemma 1.
We define the following quantities: $\bar{G}=-H^{-1}\Lambda^{\prime}\hat{\beta}_{X}$,
$\Delta_{G}=\theta_{0}\Omega_{X}$, and $\hat{\Delta}_{G}=\theta_{0}\hat{\Omega}_{X}$. }{\small\par}

{\small{}By Assumption 1, under $H_{0}:\theta=\theta_{0}$, 
\[
\begin{pmatrix}g(\theta_{0})\\
\bar{G}-\Delta_{G}\Omega^{-1}g(\theta_{0})
\end{pmatrix}\sim N\left(\begin{pmatrix}0\\
G
\end{pmatrix},\begin{pmatrix}\Omega & 0\\
0 & \Omega_{X}-\Delta_{G}\Omega^{-1}\Delta_{G}
\end{pmatrix}\right).
\]
}{\small\par}

{\small{}Let $\bar{S}_{0}=\Omega^{-\frac{1}{2}}g(\theta_{0})$ and
$\bar{T}_{0}=(\Omega_{X}-\Delta_{G}\Omega^{-1}\Delta_{G})^{-\frac{1}{2}}\bar{G}-\Delta_{G}\Omega^{-1}g(\theta_{0})$.
By the above, $\bar{S}_{0}$ and $\bar{T}_{0}$ are jointly normal
and uncorrelated, and hence are independent statistics. Let $\bar{Q}_{S,0}=\bar{S}_{0}^{\prime}\bar{S}_{0}$,
$\bar{Q}_{ST,0}=\bar{S}_{0}^{\prime}\bar{T}_{0}$, and $\bar{Q}_{T,0}=\bar{T}_{0}^{\prime}\bar{T}_{0}$.
By identical arguments used in \citet[Proof of Theorem 3.2, p.252]{Smith2007a},
conditional on ${\cal Z}_{T}\sim N\big((\Omega_{X}-\Delta_{G}\Omega^{-1}\Delta_{G})^{-\frac{1}{2}}G,I_{r}\big),$
we have that $\bar{Q}_{S,0}\overset{D}{\to}\chi^{2}(r)$, $\bar{Q}_{T,0}^{-1}\bar{Q}_{ST,0}^{2}\overset{D}{\to}\chi^{2}(1)$,
and $\big(\bar{Q}_{S,0}-\bar{Q}_{T,0}+\sqrt{(\bar{Q}_{S,0}-\bar{Q}_{T,0})^{2}+4\bar{Q}_{ST,0}^{2}}\big)\big/2\overset{D}{\to}\big(\chi^{2}(1)+\chi^{2}(r-1)-{\cal Z}_{T}^{\prime}{\cal Z}_{T}+\sqrt{(\chi^{2}(1)+\chi^{2}(r-1)-{\cal Z}_{T}^{\prime}{\cal Z}_{T})^{2}+4\chi^{2}(1)({\cal Z}_{T}^{\prime}{\cal Z}_{T})}\big)\big/2$.
We are left to show that $\bar{Q}_{S}=\bar{Q}_{S,0}+o_{P}(1)$, $\bar{Q}_{ST}=\bar{Q}_{ST,0}+o_{P}(1)$,
and $\bar{Q}_{T}=\bar{Q}_{T,0}+o_{P}(1)$, so that the result of the
theorem follows by Slutsky's lemma. }{\small\par}

{\small{}First, note that $(\hat{G}-\hat{\Delta}_{G}\hat{\Omega}(\theta_{0})^{-1}\hat{g}(\theta_{0}))-(\bar{G}-\Delta_{G}\Omega^{-1}g(\theta_{0}))=(\hat{G}-\bar{G})-(\hat{\Delta}_{G}-\Delta_{G})\hat{\Omega}(\theta_{0})^{-1}\hat{g}(\theta_{0})-\Delta_{G}(\hat{\Omega}(\theta_{0})^{-1}-\Omega^{-1})\hat{g}(\theta_{0})-\Delta_{G}\Omega^{-1}(\hat{g}(\theta_{0})-g(\theta_{0}))=(\hat{G}-\bar{G})+o_{P}(1)$,
by Lemmas 3(v) and (vi), and Lemmas 5(i) and (ii). }{\small\par}

{\small{}Using the expansion of $\hat{\lambda}_{k}-H^{-1}\lambda_{k}$
from Equation $(S.1)$, we can write 
\begin{eqnarray*}
\hat{G}-\bar{G} & = & -\sum_{k=1}^{p}(\hat{\lambda}_{k}-H^{-1}\lambda_{k})\beta_{X_{k}}-\sum_{k=1}^{p}(\hat{\lambda}_{k}-H^{-1}\lambda_{k})\varepsilon_{X_{k}}\\
 & = & H^{\prime}\frac{1}{n_{Z}}\sum_{i=1}^{n_{Z}}\sum_{k=1}^{p}F_{i}e_{ki}\beta_{X_{k}}-\Big(\frac{1}{n_{Z}}\sum_{i=1}^{n_{Z}}\hat{F}_{i}(F_{i}-H^{-1\prime}\hat{F}_{i})^{\prime}\Big)(\Lambda^{\prime}\beta_{X})\\
 &  & -\frac{1}{n_{Z}}\sum_{i=1}^{n_{Z}}(\hat{F}_{i}-H^{\prime}F_{i})\big(\sum_{k=1}^{p}e_{ki}\beta_{X_{k}}\big)-\sum_{k=1}^{p}(\hat{\lambda}_{k}-H^{-1}\lambda_{k})\varepsilon_{X_{k}}\\
 & := & {\cal R}_{4}+{\cal R}_{5}+{\cal R}_{6}+{\cal R}_{7}.
\end{eqnarray*}
By CS, $\Vert{\cal R}_{4}\Vert\leq\Vert H\Vert\cdot C_{\beta}n_{Z}^{-\frac{1}{2}}\Vert(n_{Z}p)^{-\frac{1}{2}}\sum_{i=1}^{n_{Z}}\sum_{k=1}^{p}F_{i}e_{ki}\Vert=o_{P}(1)$,
where the equality follows by Lemma 1(i), $\beta_{X_{k}}=\Theta(p^{-\frac{1}{2}})$
for all $k\in[p]$, and Assumption S0(iv). For ${\cal R}_{5}$, note
that $\Vert{\cal R}_{5}\Vert\leq\Vert H^{-1}\Vert\cdot C_{\beta}p^{\frac{1}{2}}\Vert n_{Z}^{-1}\hat{F}^{\prime}(FH-\hat{F})\Vert=O_{P}(\min(n,p)^{-1})O_{P}(p^{\frac{1}{2}})=o_{P}(1)$
by Lemma 1(ii), and Lemma B.3 of \citet[p.165]{Bai2003}, $\beta_{X_{k}}=\Theta(p^{-\frac{1}{2}})$
for all $k\in[p]$, and CS. Similarly, $\Vert{\cal R}_{6}\Vert\leq C_{\beta}\big(n_{Z}^{-1}\sum_{i=1}^{n_{Z}}\Vert\hat{F}_{i}-H^{\prime}F_{i}\Vert^{2}\big)^{\frac{1}{2}}\big(n_{Z}^{-1}\sum_{i=1}^{n_{Z}}\vert p^{-\frac{1}{2}}\sum_{k=1}^{p}e_{ki}\vert^{2}\big)^{\frac{1}{2}}=O_{P}(\min(n,p)^{-\frac{1}{2}})O_{P}(1)=o_{P}(1)$
by $\beta_{X_{k}}=\Theta(p^{-\frac{1}{2}})$ for all $k\in[p]$, Lemma
A.1 of \citet[p.159]{Bai2003}, Assumption S0(iv), M, and CS. Finally,
$\Vert{\cal R}_{7}\Vert=o_{P}(1)$ by identical arguments used in
Proof of Lemma 5(i). Hence, $(\hat{G}-\hat{\Delta}_{G}\hat{\Omega}(\theta_{0})^{-1}\hat{g}(\theta_{0}))-(\bar{G}-\Delta_{G}\Omega^{-1}g(\theta_{0}))=o_{P}(1)$. }{\small\par}

{\small{}Also, $(\hat{\Omega}_{X}-\hat{\Delta}_{G}\hat{\Omega}(\theta_{0})^{-1}\hat{\Delta}_{G})-(\Omega_{X}-\Delta_{G}\Omega^{-1}\Delta_{G})=(\hat{\Omega}_{X}-\Omega_{X})-(\hat{\Delta}_{G}-\Delta_{G})\hat{\Omega}(\theta_{0})^{-1}\hat{\Delta}_{G}-\Delta_{G}(\hat{\Omega}(\theta_{0})^{-1}-\Omega^{-1})\hat{\Delta}_{G}-\Delta_{G}\Omega^{-1}(\hat{\Delta}_{G}-\Delta_{G})=o_{P}(1)$
by Lemmas 3(i)-(vi). }{\small\par}

{\small{}Note that $\bar{T}-\bar{T}_{0}=\big((\hat{\Omega}_{X}-\hat{\Delta}_{G}\hat{\Omega}(\theta_{0})^{-1}\hat{\Delta}_{G})^{-\frac{1}{2}}-(\Omega_{X}-\Delta_{G}\Omega^{-1}\Delta_{G})^{-\frac{1}{2}}\big)(\hat{G}-\hat{\Delta}_{G}\hat{\Omega}(\theta_{0})^{-1}\hat{g}(\theta_{0}))+(\Omega_{X}-\Delta_{G}\Omega^{-1}\Delta_{G})^{-\frac{1}{2}}\big((\hat{G}-\hat{\Delta}_{G}\hat{\Omega}(\theta_{0})^{-1}\hat{g}(\theta_{0}))-(\bar{G}-\Delta_{G}\Omega^{-1}g(\theta_{0}))\big)$.
Therefore, by the above, $\bar{T}-\bar{T}_{0}=o_{P}(1)$. }{\small\par}

{\small{}Next, $\bar{S}-\bar{S}_{0}=(\hat{\Omega}(\theta_{0})^{-\frac{1}{2}}-\Omega^{-\frac{1}{2}})\hat{g}(\theta_{0})+\Omega^{-\frac{1}{2}}(\hat{g}(\theta_{0})-g(\theta_{0}))=o_{P}(1)$
by Lemmas 3(v) and (vi), and Lemmas 5(i) and (ii). }{\small\par}

{\small{}Having established the consistency of $\bar{S}$ for $\bar{S}_{0}$,
and $\bar{T}$ for $\bar{T}_{0}$, we have that $\bar{Q}_{S}-\bar{Q}_{S,0}=(\bar{S}-\bar{S}_{0})^{\prime}(\bar{S}-\bar{S}_{0})+2\bar{S}_{0}^{\prime}(\bar{S}-\bar{S}_{0})=o_{P}(1)$,
$\bar{Q}_{ST}-\bar{Q}_{ST,0}=\bar{S}^{\prime}\bar{T}-\bar{S}_{0}^{\prime}\bar{T}_{0}=(\bar{S}-\bar{S}_{0})^{\prime}(\bar{T}-\bar{T}_{0})+\bar{S}_{0}^{\prime}(\bar{T}-\bar{T}_{0})+(\bar{S}-\bar{S}_{0})^{\prime}\bar{T}_{0}=o_{P}(1)$,
and $\bar{Q}_{T}-\bar{Q}_{T,0}=(\bar{T}-\bar{T}_{0})^{\prime}(\bar{T}-\bar{T}_{0})+2\bar{T}_{0}^{\prime}(\bar{T}-\bar{T}_{0})=o_{P}(1)$. }{\small\par}

\section{Proof of Theorem 3: S-LIML conditional distribution}

{\small{}Let $\hat{G}_{{\cal S}}=\Gamma_{{\cal S}}^{\prime}\hat{G}$,
$\hat{g}_{{\cal S}}(\theta)=\Gamma_{{\cal S}}^{\prime}\hat{g}(\theta)$,
$\hat{\Omega}_{{\cal S}}(\theta)=\Gamma_{{\cal S}}^{\prime}\hat{\Omega}(\theta)\Gamma_{{\cal S}}$,
and $\hat{Q}_{{\cal S}}(\theta)=\hat{g}_{{\cal S}}(\theta)^{\prime}\hat{\Omega}_{{\cal S}}(\theta)^{-1}\hat{g}_{{\cal S}}(\theta)$.
Then, by identical arguments used in Proof of Theorem 1, $V_{{\cal S}}^{-\frac{1}{2}}(\hat{\theta}_{{\cal S}}-\theta_{0})=-V_{{\cal S}}^{\frac{1}{2}}G_{{\cal S}}^{\prime}\Omega_{{\cal S}}^{-1}g_{{\cal S}}(\theta_{0})+o_{P}(1)$,
where $V_{{\cal S}}=(G_{{\cal S}}^{\prime}\Omega_{{\cal S}}^{-1}G_{{\cal S}})^{-1}$,
$G_{{\cal S}}=\Gamma_{S}^{\prime}G$, $\Omega_{{\cal S}}=\Gamma_{{\cal S}}^{\prime}\Omega\Gamma_{{\cal S}}$,
and $g_{{\cal S}}(\theta)=\Gamma_{S}^{\prime}g(\theta)$. }{\small\par}

{\small{}Let $D$ be a $r\times r$ diagonal matrix with its $(k,k)$-th
entry given by $(\Omega_{X})_{kk}$, and $\hat{D}$ be a $r\times r$
diagonal matrix with its $(k,k)$-th entry given by $(\hat{\Omega}_{X})_{kk}$. }{\small\par}

{\small{}Also, by identical arguments used in Proof of Theorem 2,
$\hat{G}-\bar{G}=o_{P}(1)$, and $\bar{G}-G=-H^{-1}\Lambda^{\prime}\varepsilon_{X}$,
so that by T, $\hat{G}-G=-H^{-1}\Lambda^{\prime}\varepsilon_{X}+o_{P}(1)$.
Thus, by Lemma 3(ii), $D^{-\frac{1}{2}}(\hat{G}-G)=O_{P}(1)$. This
leads to the joint asymptotic normality result 
\[
\begin{pmatrix}V_{{\cal S}}^{-\frac{1}{2}}(\hat{\theta}_{{\cal S}}-\theta_{0})\\
D^{-\frac{1}{2}}(\hat{G}-G)
\end{pmatrix}\overset{D}{\to}N\left(\begin{pmatrix}0\\
0
\end{pmatrix},\begin{pmatrix}1 & \theta_{0}V_{{\cal S}}^{\frac{1}{2}}G_{{\cal S}}^{\prime}\Omega_{{\cal S}}^{-1}\Gamma_{{\cal S}}^{\prime}\Omega_{X}D^{-\frac{1}{2}}\\
D^{-\frac{1}{2}}\Omega_{X}\Gamma_{{\cal S}}\Omega_{{\cal S}}^{-1}G_{{\cal S}}V_{{\cal S}}^{\frac{1}{2}}\theta_{0} & D^{-\frac{1}{2}}\Omega_{X}D^{-\frac{1}{2}}
\end{pmatrix}\right).
\]
}{\small\par}

{\small{}By Lemmas 3(vi) and 4(i), we have $(\hat{D}^{-\frac{1}{2}}-D^{-\frac{1}{2}})(\hat{G}-G)=o_{P}(1)$.
Also, under ${\cal H}_{0j}:G{}_{j}=0$, we have $(\hat{D}^{-\frac{1}{2}}G)_{j}=0$.
Note that $\hat{{\cal T}}=({\cal T}_{1},\ldots,{\cal T}_{r})^{\prime}=\hat{D}^{-\frac{1}{2}}\hat{G}$.
Using the above normality result, approximately, 
\[
\begin{pmatrix}\hat{\theta}_{{\cal S}}-\theta_{0}\\
\hat{{\cal T}}
\end{pmatrix}\overset{a}{\sim}N\left(\begin{pmatrix}0\\
\hat{D}^{-\frac{1}{2}}G
\end{pmatrix},\begin{pmatrix}V_{{\cal S}} & C_{G}^{\prime}\\
C_{G} & V_{G}
\end{pmatrix}\right),
\]
where $C_{G}=-D^{-\frac{1}{2}}\Omega_{X}\Gamma_{{\cal S}}\Omega_{{\cal S}}^{-1}G_{{\cal S}}V_{{\cal S}}\theta_{0}$
and $V_{G}=D^{-\frac{1}{2}}\Omega_{X}D^{-\frac{1}{2}}$. }{\small\par}

{\small{}To remove dependence of $\hat{\theta}_{{\cal S}}$ on $\hat{{\cal T}}$,
we condition on a sufficient statistic for the unknown nuisance parameter
$\hat{D}^{-\frac{1}{2}}G$ which, under $H_{0}:\theta=\theta_{0}$,
is given by $U=\hat{{\cal T}}-C_{G}V_{{\cal S}}^{-1}(\hat{\theta}_{{\cal S}}-\theta_{0})$.
We can also write $\hat{\theta}_{{\cal S}}-\theta_{0}=V_{{\cal S}}^{\frac{1}{2}}{\cal K}+o_{P}(n^{-\frac{1}{2}})$,
where ${\cal K}\sim N(0,1)$. Thus, if ${\cal S}$ is the selection
event, and ${\cal R}$ is the set of indices of the estimated factor
loadings corresponding to the retained instruments, approximately,
\[
\mathbb{P}(\hat{\theta}_{{\cal S}}\leq w+\theta_{0}\vert{\cal S},U=u)\overset{a}{\sim}\frac{\mathbb{P}_{m}\big(\{V_{{\cal S}}^{\frac{1}{2}}{\cal K}\leq w\}\bigcap_{j\in{\cal R}}\{\vert(\bar{u})_{j}\vert>c_{\delta}\}\bigcap_{k\in[r]\backslash{\cal R}}\{\vert(\bar{u})_{k}\vert\leq c_{\delta}\}\big)}{\mathbb{P}_{m}\big(\bigcap_{j\in{\cal R}}\{\vert(\bar{u})_{j}\vert>c_{\delta}\}\bigcap_{k\in[r]\backslash{\cal R}}\{\vert(\bar{u})_{k}\vert\leq c_{\delta}\}\big)},
\]
for $m\to\infty$, where $\bar{u}=u+C_{G}V_{{\cal S}}^{-\frac{1}{2}}{\cal K}$,
and $\mathbb{P}_{m}(.)$ denotes the empirical distribution based
on $m$ draws of ${\cal K}\sim N(0,1)$. In practice, we also replace
the variance components $V_{{\cal S}}$, $C_{G}$ and $V_{G}$ with
consistent estimates $\hat{V}_{{\cal S}}=(\hat{G}_{{\cal S}}^{\prime}\hat{\Omega}_{{\cal S}}^{-1}\hat{G}_{{\cal S}})^{-1}$,
$\hat{C}_{G}=-\hat{D}^{-\frac{1}{2}}\hat{\Omega}_{X}\Gamma_{{\cal S}}\hat{\Omega}_{{\cal S}}^{-1}\hat{G}_{{\cal S}}\hat{V}_{{\cal S}}\hat{\theta}_{{\cal S}}$,
and $\hat{V}_{G}=\hat{D}^{-\frac{1}{2}}\hat{\Omega}_{X}\hat{D}^{-\frac{1}{2}}$. }{\small\par}

\section{Justification of covariance structure in Assumption 1}

{\small{}Under Assumption 1, for any variants $j,k\in[p]$, $Cov(\hat{\beta}_{X_{j}},\hat{\beta}_{X_{k}})=(\Sigma_{X})_{jk}=\rho_{jk}\sigma_{X_{j}}\sigma_{X_{k}}$
and $Cov(\hat{\beta}_{Y_{j}},\hat{\beta}_{Y_{k}})=(\Sigma_{Y})_{jk}=\rho_{jk}\sigma_{Y_{j}}\sigma_{Y_{k}}$.
We justify this for our real data application which involved summary
data reported using: (i) a linear model for a continuous exposure:
$\mathbb{E}[x\vert z_{j}]=\alpha_{X_{j}}+\beta_{X_{j}}z_{j}$; (ii)
a logit model for a binary outcome: $\mathbb{E}[y\vert z_{j}]=L(\alpha_{Y_{j}}+\beta_{Y_{j}}z_{j})$,
where $L$ is the logit function where, for any $a\in\mathbb{R}$,
$L(a)=(1+\exp(-a))^{-1}$. }{\small\par}

\subsection{Continuous exposure}

{\small{}We describe genetic associations as `weak' if they decrease
with the sample size at some rate. Suppose $\beta_{X_{j}}=O(n_{X}^{-\kappa})$
for any $\kappa>0$. Then, for any variant $j$, the residual variance
of $x$ is given by $Var(x)-\beta_{X_{j}}^{2}Var(z_{j})=Var(x)+O(n_{X}^{-2\kappa})$,
and so in the linear model, the variance of the genetic association
satisfies $(\Sigma_{X})_{jj}=n_{X}^{-1}Var(z_{j})^{-1}Var(x)+O(n_{X}^{-(1+2\kappa)})$.
Similarly, the covariance of residual errors of the $j$-th and $k$-th
variant associations satisfies $Var(x)-\beta_{X_{j}}^{2}Var(z_{j})-\beta_{X_{k}}^{2}Var(z_{k})+\beta_{X_{j}}\beta_{X_{k}}Var(z_{j})^{\frac{1}{2}}Var(z_{k})^{\frac{1}{2}}\rho_{jk}=Var(x)+O(n_{X}^{-2\kappa})$,
and so $(\Sigma_{X})_{jk}=n_{X}^{-1}Var(z_{j})^{-\frac{1}{2}}Var(z_{k})^{-\frac{1}{2}}\rho_{jk}Var(x)+O(n_{X}^{-(1+2\kappa)})$.
Therefore, the covariance term $(\Sigma_{X})_{jk}$ satisfies $(\Sigma_{X})_{jk}=\sigma_{X_{j}}\sigma_{X_{k}}\rho_{jk}+O(n_{X}^{-(1+\kappa)})$.
Since $\sigma_{X_{j}}\sigma_{X_{k}}\rho_{jk}=\Theta(n_{X}^{-1})$,
the $o(n_{X}^{-1})$ remainder term is negligible. }{\small\par}

\subsection{Binary outcome}

{\small{}Under weak variant--outcome associations, $\beta_{Y_{j}}=O(n_{Y}^{-\kappa})$
for some $\kappa>0$. For any $a\in\mathbb{R}$, let $L_{1}(a)=(1+\exp(-a))^{-2}\exp(-a)$,
and $L_{2}(a)=(1+\exp(-a))^{-3}\exp(-a)(1-\exp(-a))$. }{\small\par}

{\small{}For any variant $j$, by the mean value theorem, since $\vert z_{j}\vert=O(1)$
and $\beta_{Y_{j}}=O(n_{Y}^{-\kappa})$, for some $\bar{\beta}_{Y_{j}}\in(0,\beta_{Y_{j}})$,
\begin{eqnarray*}
L(\alpha_{Y_{j}}+\beta_{Y_{j}}z_{j}) & = & L(\alpha_{Y_{j}})+L_{1}(\alpha_{Y_{j}}+\bar{\beta}_{Y_{j}}z_{j})\beta_{Y_{j}}z_{j}\\
 & = & L(\alpha_{Y_{j}})+O(1)O(n_{Y}^{-\kappa})\\
 & = & L(\alpha_{Y_{j}})+o(1),
\end{eqnarray*}
and similarly, $L_{1}(\alpha_{Y_{j}}+\beta_{Y_{j}}z_{j})=L_{1}(\alpha_{Y_{j}})+L_{2}(\alpha_{Y_{j}}+\bar{\beta}_{Y_{j}}z_{j})\beta_{Y_{j}}z_{j}=L_{1}(\alpha_{Y_{j}})+o(1)$.
Since the genetic effects are weak, for each variant $j$, as $n_{Y}\to\infty$,
the probability $\mathbb{P}(y=1\vert z_{j})$ is not dependent on
$z_{j}$, and the constants $\alpha_{Y_{j}}$ converge to $\alpha_{0}$,
for some constant $\alpha_{0}$. Let $L=L(\alpha_{0})$ and $L_{1}=L_{1}(\alpha_{0})$.
Also, let $w_{j}=(1,z_{j})^{\prime}$ and $\gamma_{j}=(\alpha_{Y_{j}},\beta_{Y_{j}})^{\prime}$. }{\small\par}

{\small{}If $\hat{\gamma}_{j}=(\hat{\alpha}_{Y_{j}},\hat{\beta}_{Y_{j}})^{\prime}$
is the logit regression estimate for the model $\mathbb{P}(y=1\vert z_{j})=L(w_{j}^{\prime}\gamma_{j})$,
by standard asymptotic arguments, 
\[
\hat{\gamma}_{j}-\gamma_{j}\overset{a}{=}\mathbb{E}[w_{j}w_{j}^{\prime}L_{1}]^{-1}\frac{1}{n_{Y}}\sum_{i=1}^{n_{Y}}w_{ji}(y_{i}-L(w_{ji}^{\prime}\gamma_{j}))\overset{a}{\sim}N(0,\sigma_{\gamma_{j}}^{2}),
\]
where $\sigma_{\gamma_{j}}^{2}=n_{Y}^{-1}\mathbb{E}[w_{j}w_{j}^{\prime}L_{1}]^{-1}$.
Note that the asymptotic variance $(\Sigma_{Y})_{jj}$ of the estimated
slope coefficient $\hat{\beta}_{Y_{j}}$, corresponds to the $(2,2)$-th
element of the covariance matrix $\sigma_{\gamma_{j}}^{2}$. In particular,
$(\Sigma_{Y})_{jj}:=\sigma_{Y_{j}}^{2}=n_{Y}^{-1}L_{1}^{-1}Var(Z_{j})^{-1}$.
Furthermore, for any two variants $j,k$, $\sqrt{n_{Y}}(\hat{\beta}_{Y_{j}}-\beta_{Y_{j}})$
and $\sqrt{n_{Y}}(\hat{\beta}_{Y_{k}}-\beta_{Y_{k}})$ are asymptotically
jointly normal with asymptotic covariance 
\[
n_{Y}(\Sigma_{Y})_{jk}:=\mathbb{E}[w_{j}w_{j}^{\prime}L_{1}]^{-1}\mathbb{E}[w_{j}w_{k}^{\prime}(y-L)^{2}]\mathbb{E}[w_{k}w_{k}^{\prime}L_{1}]^{-1\prime}.
\]
}{\small\par}

{\small{}Note that 
\begin{eqnarray*}
\mathbb{E}[w_{j}w_{k}^{\prime}(y-L)^{2}] & = & (1-L)^{2}\mathbb{E}[w_{j}w_{k}^{\prime}]\mathbb{P}(y=1)+L^{2}\mathbb{E}[w_{j}w_{k}^{\prime}]\mathbb{P}(y=0)+o(1)\\
 & = & L(1-L)\mathbb{E}[w_{j}w_{k}^{\prime}]+o(1),
\end{eqnarray*}
since, as $n_{Y}\to\infty$, $\mathbb{P}(y=1)=\sum_{l=0}^{2}\mathbb{P}(y=1\vert z_{j})\mathbb{P}(z_{j}=l)+o(1)=L\sum_{l=0}^{2}\mathbb{P}(z_{j}=l)+o(1)=L+o(1)$,
and therefore $\mathbb{P}(y=0)=1-L+o(1)$. }{\small\par}

{\small{}Overall, 
\begin{eqnarray*}
n_{Y}(\Sigma_{Y})_{jk} & = & \frac{L(1-L)}{L_{1}^{2}}\mathbb{E}[w_{j}w_{j}^{\prime}]^{-1}\mathbb{E}[w_{j}w_{k}^{\prime}]\mathbb{E}[w_{k}w_{k}^{\prime}]^{-1\prime}\\
 & = & \frac{L(1-L)}{L_{1}^{2}}\frac{1}{Var(z_{j})Var(z_{k})}\begin{bmatrix}\mathbb{E}[z_{j}^{2}] & -\mathbb{E}[z_{j}]\\
-\mathbb{E}[z_{j}] & 1
\end{bmatrix}\begin{bmatrix}1 & \mathbb{E}[z_{k}]\\
\mathbb{E}[z_{j}] & \mathbb{E}[z_{j}z_{k}]
\end{bmatrix}\begin{bmatrix}\mathbb{E}[z_{k}^{2}] & -\mathbb{E}[z_{k}]\\
-\mathbb{E}[z_{k}] & 1
\end{bmatrix}\\
 & = & \frac{L(1-L)}{L_{1}^{2}}\frac{1}{Var(z_{j})Var(z_{k})}\begin{bmatrix}Var(z_{j}) & \mathbb{E}[z_{j}^{2}]\mathbb{E}[z_{k}]-\mathbb{E}[z_{j}]\mathbb{E}[z_{j}z_{k}]\\
0 & Cov(z_{j},z_{k})
\end{bmatrix}\begin{bmatrix}\mathbb{E}[z_{k}^{2}] & -\mathbb{E}[z_{k}]\\
-\mathbb{E}[z_{k}] & 1
\end{bmatrix}\\
 & = & \frac{1}{L_{1}Var(z_{j})Var(z_{k})}\begin{bmatrix}\bar{\sigma}_{jk}^{2} & -\mathbb{E}[z_{j}]Cov(z_{j},z_{k})\\
-\mathbb{E}[z_{k}]Cov(z_{j},z_{k}) & Cov(z_{j},z_{k})
\end{bmatrix},
\end{eqnarray*}
where $\bar{\sigma}_{jk}^{2}=\mathbb{E}[z_{j}^{2}]\mathbb{E}[z_{k}^{2}]-\mathbb{E}[z_{j}]^{2}\mathbb{E}[z_{k}^{2}]-\mathbb{E}[z_{j}^{2}]\mathbb{E}[z_{k}]^{2}+\mathbb{E}[z_{j}]\mathbb{E}[z_{k}]\mathbb{E}[z_{j}z_{k}]$,
and the last line follows by noting that $L(1-L)=L_{1}$. }{\small\par}

{\small{}Thus, the covariance of $\hat{\beta}_{Y_{j}}$ and $\hat{\beta}_{Y_{k}}$
is approximately given by 
\begin{eqnarray*}
(\Sigma_{Y})_{jk} & = & \frac{Cov(z_{j},z_{k})}{n_{Y}L_{1}Var(z_{j})Var(z_{k})}\\
 & = & \frac{1}{\sqrt{n_{Y}L_{1}Var(Z_{j})}}\cdot\frac{1}{\sqrt{n_{Y}L_{1}Var(Z_{k})}}\cdot\frac{Cov(z_{j},z_{k})}{\sqrt{Var(z_{j})Var(z_{k})}}\\
 & = & \sigma_{Y_{j}}\sigma_{Y_{k}}\rho_{jk},
\end{eqnarray*}
where $\rho_{jk}$ is the correlation between $z_{j}$ and $z_{k}$. }{\small\par}

\section{Simulation results: pruned CLR with invalid IVs}
\begin{center}
{\small{}\includegraphics[width=16.5cm]{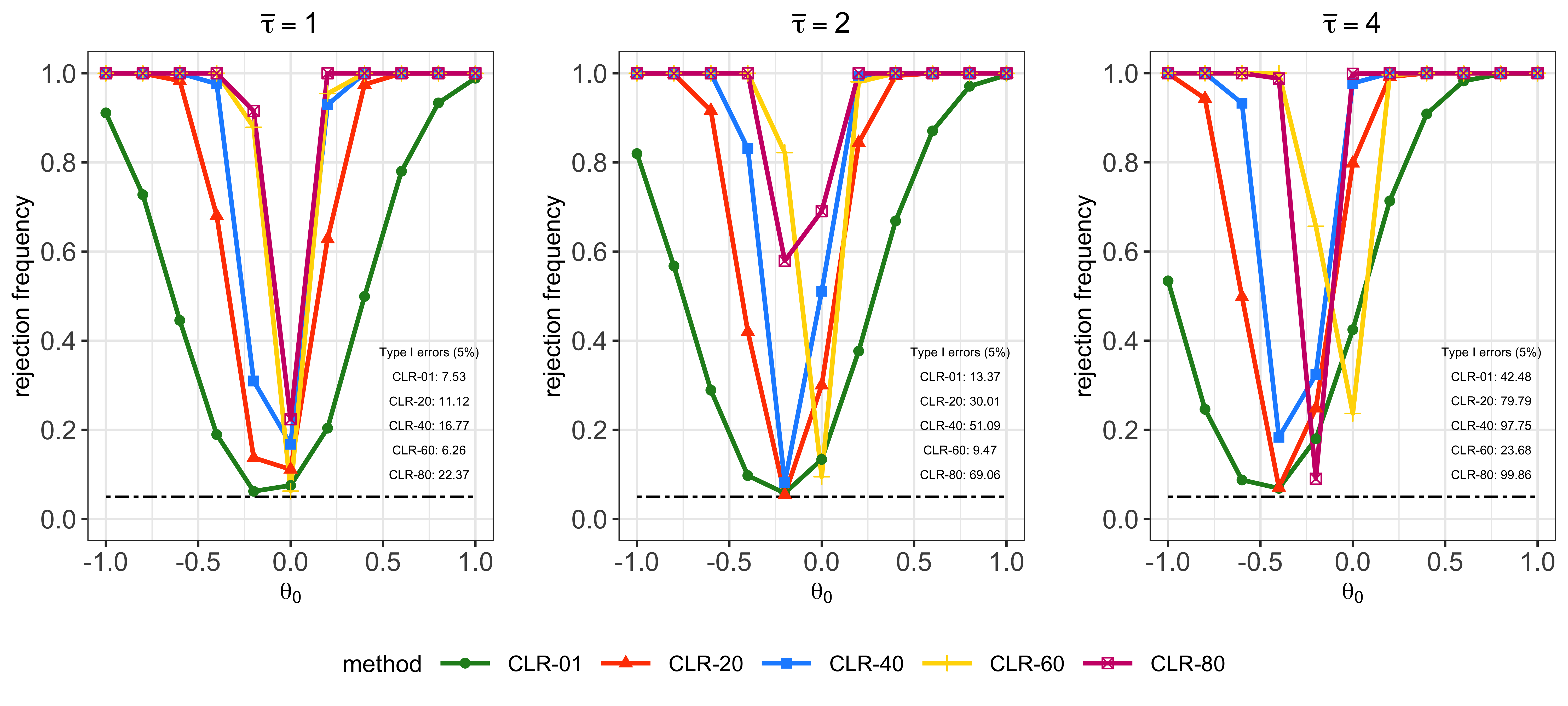}}{\footnotesize{}}\\
{\footnotesize{}Figure S.1. Sensitivity to pruning thresholds: inflated
type I errors with invalid instruments when testing the null hypothesis
${\cal H}_{0}:\theta_{0}=0$. }{\footnotesize\par}
\par\end{center}

{\small{}For the simulation design of Section 4.2 in the main text,
we had presented simulation results for the CLR test using individual
variants as instruments up to the correlation threshold $R^{2}=0.6$.
Here we present the results which include other pruning thresholds
that we had considered. In particular, Figure S.1 highlights that
under invalid instruments the CLR test can be quite sensitive to the
pruning threshold chosen. On the one hand, we would like to include
more variants as instruments to `even out' the invalid instruments
problem under this particular case of balanced direct variant effects.
However, the substantially inflated type I error rates under the $R^{2}=0.8$
threshold suggest that including highly correlated variants may also
lead to unreliable tests. Overall, we suggest it may be difficult
in practice to determine a suitable pruning threshold for robust and
precise inferences. }{\small\par}
\end{document}